\documentclass[aps,pre,onecolumn,showkeys,superscriptaddress,nofootinbib,nobibnotes,10pt]{revtex4-2}
\usepackage[utf8]{inputenc}
\usepackage[T1]{fontenc}
\usepackage{graphicx}
\usepackage{amsmath,amssymb}
\usepackage{mathptmx}
\usepackage{braket}
\usepackage{comment}
\usepackage{lineno}
\usepackage[hidelinks]{hyperref}
\providecommand{\widebar}[1]{\overline{#1}}
\renewcommand{\doi}[1]{\href{https://doi.org/#1}{\nolinkurl{#1}}}

\hypersetup{pdftitle={Discrete time crystals and heteroclinic switching in actively mode-locked semiconductor lasers},pdfauthor={Elias R. Koch; Albert Gallemí; Clàudia López; Finn Petersen; Julien Javaloyes; Svetlana V. Gurevich}}

\begin{document}
\title{Discrete time crystals and heteroclinic switching in actively mode-locked semiconductor lasers}
\author{Elias R. Koch}
\affiliation{Institute for Theoretical Physics, University of Münster, Wilhelm-Klemm-Str. 9, 48149, Münster, Germany}
\author{Albert Gallemí}
\affiliation{Departament de Física, Universitat de les Illes Balears, \& Institute of Applied Computing and Community Code (IAC-3), C/ Valldemossa km 7.5, 07122, Mallorca, Spain}
\author{Clàudia López}
\affiliation{Institute for Theoretical Physics, University of Münster, Wilhelm-Klemm-Str. 9, 48149, Münster, Germany}
\author{Finn Petersen}
\affiliation{Institute for Theoretical Physics, University of Münster, Wilhelm-Klemm-Str. 9, 48149, Münster, Germany}
\author{Julien Javaloyes}
\email{Julien.Javaloyes@uib.es}
\affiliation{Departament de Física, Universitat de les Illes Balears, \& Institute of Applied Computing and Community Code (IAC-3), C/ Valldemossa km 7.5, 07122, Mallorca, Spain}
\author{Svetlana V. Gurevich}
\affiliation{Institute for Theoretical Physics, University of Münster, Wilhelm-Klemm-Str. 9, 48149, Münster, Germany}
\affiliation{Center for Data Science and Complexity, University of Münster, Corrensstr. 2, 48149, Münster, Germany}
\date{}

\begin{abstract}
We investigate the mechanisms underlying the emergence of discrete time crystals in actively mode-locked semiconductor lasers. Our analysis is based on a Haus master equation generalized to retain carrier memory between round-trips. We perform a detailed bifurcation analysis that reveals a well-defined hierarchy of subharmonic states, illustrating how pulses may suppress their neighbors when the modulation runs faster than the pulse round-trip frequency. To identify the underlying mechanism, we derive effective equations of motion that couple the pulse energies, their positions relative to the modulation transmission maxima, and the local time-dependent gain. These reveal that transitions between time-crystalline states occur via transcritical global bifurcations, opening a regime of heteroclinic winnerless competition between symmetry-degenerate configurations.
\end{abstract}
\keywords{Discrete time crystals, Active mode-locking, Semiconductor lasers, Bifurcations, Heteroclinic cycles, Winnerless competition}
\maketitle

\section{Introduction}

Engineering and shaping light play a central role in photonics. Mode-locked (ML) lasers~\cite{H_JSTQE_00} provide direct access to highly coherent pulse trains and optical frequency combs, combining ultrashort pulses with precisely defined spectral mode spacing. These properties enable a broad range of applications, including precision spectroscopy, frequency metrology, astronomy, and atmospheric sensing~\cite{HUH_PRL_00,CNS_O_16,UHH_NAT_02,MUH_RAS_07,D_JOSAB_10,PPR_PR_18}.
Although ML is a mature laser technology, it remains a key platform for both photonic applications and fundamental studies of complex collective dynamics~\cite{K-Nature-03,WSP-Nature-20,LBY-RMP-15,CZG-Optica-24,SBG-PRL-25}.

\begin{figure}[htbp]
    \centering
    \includegraphics[width=0.7\textwidth]{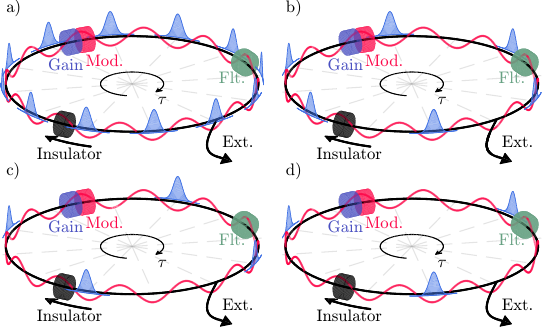}
    \caption{(a) A schematic of a unidirectional actively mode-locked ring cavity containing twelve pulses which are situated around the transmission minima imposed by the modulation (red line). The different components for gain (blue), modulator (red), filter (green) and insulator (black) are indicated as small colored cylinders. A small arrow illustrates the extraction of intensity. The round-trip time of the cavity is $\tau$. (b-d) Subharmonic TC regimes, where the conventional HML$_{12}$ state of (a) is replaced by a state possessing a period that is an integer multiple of the driving period.}\label{fig:AML_sketch}
\end{figure}

In particular, \emph{actively mode-locked} (AML) lasers employ intracavity acousto-optic or electro-optic modulators driven by radio-frequency signals to generate highly coherent ultrashort pulses~\cite{SIEGMAN-BOOK,KS2_JQE_70,TK-JQE-09,CDDC_OL_09,KKL_PhRes_17,KKWK_OL_94,BWX_AFM_18}. Their electrical tunability allows precise control of the repetition rate, enabling applications in high-resolution spectroscopy, waveform generation, microwave photonics, and broadband optical communications~\cite{KS2_JQE_70,BWX_AFM_18,QOG-JOA-09,CW-NAP-10,WHZ-Light-25}. Long-cavity AML lasers can operate in the \emph{harmonic mode-locking} (HML) regime, where multiple equidistant pulses circulate in the cavity, effectively multiplying the repetition rate~\cite{QOG-JOA-09,WHZ-Light-25}, see Fig.~\ref{fig:AML_sketch}~(a) for the schematic. High harmonic numbers lead to a system
with many degrees of freedom, and a variety of techniques have been developed to regularize pulse trains and improve their coherence~\cite{YKN-APL-92,HM-OL-93,SS-EL-93,DHI-OL-94,CD-OL-96,TGK-OL-00}. However, unstable regimes and dropout phenomena in systems with slow gain media have been reported~\cite{ZPH-JQE-05,ZH-JLT-06,LSB-OECC-07}. In contrast, the
fast gain in semiconductors promotes strong pulse interactions that regularize the dynamics~\cite{KCB-JQE-98,JCM-PRL-16}. 

Since AML lasers cannot directly produce arbitrary pulse sequences, they typically require additional complex pulse-picking mechanisms. Recently, new subharmonic regimes have been observed in a unidirectional ring AML semiconductor laser operating at room temperature~\cite{WKY_PRL_26} where alternate pulses vanish, linking AML dynamics to \emph{discrete time-crystal} (TC) phases, see Fig.~\ref{fig:AML_sketch}~(b-d).

In general, continuous~\cite{MIQ-N-22,AHM-NAM-21,LOM-NP-23,CCK-SCI-24} and discrete~\cite{S-PRA-15,EBN-PRL-16,KLM-PRL-16,ZHK-N-17,BTJ-NC-19} TCs appear in systems where the continuous or discrete time-translation symmetry imposed by their drive is broken. In particular, discrete TCs can emerge in periodically driven platforms ranging from trapped atomic ions, driven disordered spin ensembles in diamond and superconducting qubit systems to optical Kerr cavities, to name just a few~\cite{KKP-PRL-21,CCL-N-17, FR-SA-22, TMM_NC_22}. In the AML setting reported in Ref.~\cite{WKY_PRL_26}, discrete TC phases appear when the bias current or the modulation frequency is slightly blue detuned with respect to the HML resonance, replacing the conventional HML state with a state possessing a period that is an integer multiple of the driving period, cf. Fig.~\ref{fig:AML_sketch}~(b-d).

Beyond their fundamental novelty, these states open entirely new avenues for controlling pulse repetition rates in AML lasers. Traditionally, the pulse width and repetition rate are strongly coupled and, following the celebrated Kuizenga-Siegman theory \cite{KS_JQE_70}, driving the modulator at longer periods produces shallower effective potentials, leading to broader pulses. By contrast, discrete TC states provide a robust mechanism for decoupling pulse width from repetition rate. This independent control not only represents a conceptual leap in breaking discrete time-translation symmetry but also offers powerful tools for applications in precision metrology, optical frequency comb engineering, and photonic information processing. The experimental results reported in Ref.~\cite{WKY_PRL_26} were confirmed by a theoretical time-delayed model adapted to the large gain and losses, which are typical for semiconductor AML systems~\cite{KGJ-OL-24}. However, the emergence mechanisms of the TC states in the AML system are still not fully understood. One reason is that the numerical analysis of the \emph{delay differential equation} (DDE) model studied in Ref.~\cite{KGJ-OL-24} requires the continuation of periodic and quasiperiodic orbits with long periods, making bifurcation analysis tedious.

This motivates the development of complementary effective models, such as the widely used \emph{Haus master equation} (HME), which constitutes a \emph{partial differential equation} (PDE) model for ML~\cite{H_IEE_75,H_JSTQE_00}. The HME is usually derived under the assumptions of small gain, weak losses, and weak spectral filtering, while restricting the field dynamics to a short temporal interval around the pulse. In particular, the HME was originally derived for gain media that evolve slowly on the timescale of the cavity round trip, resulting in a nearly uniform temporal gain profile within the cavity. How memory effects mediated by the incomplete recovery of the gain influence pulse dynamics in intermediate cavity regimes has remained an open question. This is of particular interest in semiconductor lasers, where gain recovery occurs on a nanosecond timescale. Recently, several generalizations of the HME preserving carrier memory from one round trip to the next were introduced~\cite{HLGJ_OL_20,PGG-NAC-20,NV_PRE_21}, allowing for the description of complex pulse dynamics, including Q-switched transitions and HML regimes. However, since these suggested generalizations involved non-local structures or non-instantaneous boundary conditions, no bifurcation analysis has been conducted so far.

In this paper, we systematically analyze multipulse dynamics in a semiconductor AML laser and identify the mechanisms underlying the emergence of discrete TCs. Employing a detailed nonlinear analysis of an extended HME for AML, we map a well-defined hierarchy of distinct subharmonic TC states organized in a fixed order and identify the bifurcation scenario driving the transitions between them. To gain further insight into the mechanisms underlying TC formation, we derive effective \emph{equations of motion} (EOM) capturing the essential physics of AML systems by explicitly coupling the pulse amplitudes, their alignment relative to the transmission maxima, and local gain depletion.
Beyond providing a particularly intuitive physical explanation for the emergence of time crystals, we demonstrate that, from the point of view of nonlinear dynamics, the transition between different TC states appears as a global transition that emerges from a collision with a transcritical bifurcation~\cite{CSA_ScRep_18}. This bifurcation ultimately paves the way for a regime of heteroclinic cycles that mediate a \emph{winnerless competition}~\cite{HR_JMAA_09} between symmetry-degenerate TC configurations.

\section{Haus master equation}

We start by considering the Haus master PDE~\cite{H_IEE_75,H_JSTQE_00}. Besides leading more directly to analytical results, such iterative pulse mappings are also computationally efficient: while the HML pulses are periodic solutions of, e.g., delay differential equation models~\cite{VT-PRA-05,KGJ-OL-24}, they become steady states of a one-dimensional PDE, which facilitates further bifurcation analysis. For instance, the branches of periodic solutions of PDEs can in principle be computed using numerical path-continuation while the quasi-periodic solutions of DDEs cannot be easily evaluated with DDE path-continuation tools - another argument that makes the PDE approach appealing.

To retain the long-term carrier dynamics, we adapt the approach of Ref.~\cite{HLGJ_OL_20}, initially developed for passively ML systems, to AML lasers. The HME for the slow evolution of the electric field $E=E(t,\theta)$ from one round trip to the next and the variation of the carrier density $G=G(t,\theta)$ on the fast scale $t\in(0,\,\tau)$ can be derived from, e.g., the DDE model for active mode-locking~\cite{KGJ_OL_24} using a two-timescale approach in the uniform field limit. This leads to
\begin{eqnarray}
	\partial_{\theta}E&=&\left(\frac{1-i\alpha}{2}G-\frac{k}{2}+\frac{1}{2\gamma^2}\partial_t^2+\upsilon\partial_t+m\cos(\omega_m\,t)\right)E, \label{eq:HME1} \\
	\partial_tG&=&\Gamma(G_0-G-G|E|^2)\,. \label{eq:HME2}
\end{eqnarray}
Here, $\Gamma$ is the carrier relaxation rate, $\gamma$ is the gain bandwidth, $G_0$ is the pumping rate, $\alpha$ is the linewidth enhancement factor and $k$ represents the cavity losses. Further, $m$ is the amplitude of the active modulation. We choose $m\in \mathbb{R}$ which corresponds to the case of an intensity modulator. The $N$-th harmonic mode-locked tongue creates $N$ slots for the pulses within the round-trip so that $\omega_m=2\pi N/\tau$. As in Ref.~\cite{KZM-PRL-99}, small deviations from this resonance condition are represented by the residual drift operator $\upsilon\partial_t$ applied to the field. As we operate in the reference frame of the modulator, too fast a modulation with respect to the cavity round-trip (blue detuning) corresponds to an effective motion of the pulse to the right and to $\upsilon<0$.

Although the boundary conditions for the electric field $E$ can be chosen as periodic, we use an \emph{asynchronous boundary condition} for the population inversion $G$, as in Ref.~\cite{HLGJ_OL_20}. This condition expresses the temporal continuity of the solution in a two-timescale representation, i.e., $G(\tau,\theta)=G(0,\theta+1)$. Assuming a slow evolution of the gain over the slow timescale leads to an ordinary differential equation for $g(\theta)=G(0,\theta)$, i.e., the gain value at the beginning of each round trip, that reads
\begin{equation}
	\dfrac{dg}{d\theta}=G(\tau,\theta)-g. \label{eq:BC_HME}
\end{equation}
The gain at the end of the domain, $G(\tau,\theta)$ drives the evolution of $g(\theta)$ over the slow time scale $\theta$. In turn, $g(\theta)$ provides the initial condition for solving the carrier evolution over the fast time scale, i.e., $G(0,\theta)=g(\theta)$. It was shown in Ref.~\cite{HLGJ_OL_20} that this approach can properly reproduce the Q-switch regimes and the harmonic transitions in the case of passive mode-locking.

The lasing threshold $G_{\mathrm{th}}$ is determined by the value of $G_0$ where the off-state of HME Eqs.~\eqref{eq:HME1}-\eqref{eq:BC_HME} $(E,G) = (0,G_0)$ becomes linearly unstable. The associated linear problem for the field amounts to setting $G=G_0$ in Eq.~\eqref{eq:HME1} and finding field solutions with zero growth rate.
For pulsating solutions, the transmission maxima are located at $t_j=jT$ with $T=2\pi/\omega_m$. Separating the linear contributions of gain and losses, assuming zero detuning, and expanding the modulation around one of these maxima leads to the \emph{Sturm-Liouville problem} in the parabolic approximation, which is appropriate for narrow pulses when $\gamma\gg 1$: 
\begin{eqnarray}
\left(\frac{1}{2\gamma^{2}}\partial_{t}^{2}-\frac{m\omega_m^{2}}{2}t^{2}\right)h_{j}\left(t\right) & = & -\frac{\lambda_{j}}{2}h_{j}\left(t\right)\label{eq:SL}
\end{eqnarray}
where the $h_{j}$ are the Gauss-Hermite functions and $\lambda_{j}$ is the associated eigenvalue. For $\upsilon\ll 1$ the eigenfunction of Eq.~\eqref{eq:SL} with the smallest eigenvalue, and thereby the smallest lasing threshold, is a shifted Gaussian function given by
\begin{equation}
 h_0(t)=\left(\frac{2\beta}{\pi}\right)^\frac{1}{4}e^{-\beta (t-a)^2}\,,
 \label{eq:HGM0}
\end{equation}
where a normalization constant is chosen such that the associated intensity is unity and $\beta=\sqrt{m}\gamma\omega_m/2$, while the position offset reads $a=-\gamma\upsilon/(\sqrt{m}\omega_m)$ (see Appendix~\ref{app:derivation} for the details).
We find that the corresponding eigenvalue is $\lambda_0(\upsilon)=\sqrt{m}\omega_m/\gamma+(\gamma\upsilon)^2$. The latter corresponds to the additional loss incurred by a pulsating regime as compared with CW emission. 
As predicted by the Kuizenga-Siegman theory \cite{KS_JQE_70}, the pulse width scales as $\sim 1/\sqrt{\beta}=m^{-1/4}/\sqrt{\gamma\omega_m}$ and a broader filter reduces the losses while a faster modulation increases them. The deviation from $\upsilon=0$ also induces extra losses since the pulse occupies an equilibrium position $a$ that is shifted from the maxima of transmission. Finally, for pulses narrow compared to the modulator period, the lasing threshold can be approximated as
\begin{equation}
G_0^{\textrm{th}}=k-2m+\frac{\sqrt{m}\omega_m}{\gamma}+(\gamma\upsilon)^2 \label{eq:Gth_0}
\end{equation}
which is a parabolic function in $\upsilon$ (see Appendix~\ref{app:derivation} for the detailed derivations). For large detuning values, $\upsilon\gg 1$, one expects small-amplitude intensity oscillations that are no longer localized in a sequence of pulses. In that case, the parabolic approximation does not hold but we find at leading order in $1/\upsilon$ that there is a single eigenfunction with eigenvalue $\lambda=2m$ that reads $h_{\infty}=\exp\left[-\frac{m}{\omega_m\upsilon}\sin\left(\omega_m t\right)\right]$. The lasing threshold of this modulated solution converges to that of the cavity without a modulator, i.e., $G_0^{\infty} = k$.

\subsection{Bifurcation analysis of the Haus master equation}

We operate the AML system given by Eqs.~\eqref{eq:HME1}-\eqref{eq:BC_HME} in the regime of discrete TCs~\cite{WKY_PRL_26}. 
To this end, we start our analysis by setting $t\in(0,\,\tau)$ with $\tau=4$ and introducing $N=4$ slots of the modulation leading to $\omega_m=2\pi$. Our parameters are chosen to correspond to the case of a semiconductor laser cavity whose round-trip time is $\sim 400\,$ps and that is modulated around 10~GHz, corresponding to its fourth resonance. 

We note, however, that the experiment presented in Ref.~\cite{WKY_PRL_26} used a cavity whose round-trip time was $\sim 54\,$ns, more than two orders of magnitude longer than in the study at hand, and that this cavity was modulated around its 72nd resonance. The study of the bifurcation structure of such a complex system is not achievable by numerical continuation, but the smaller subsystem presented below already exhibits a very rich and informative bifurcation structure.

We follow the stationary solutions of Eqs.~\eqref{eq:HME1}-\eqref{eq:BC_HME} in parameter space, employing pseudo-arclength continuation within the \emph{pde2path} framework~\cite{Ueker-SIAM-21}.
The steady-state solutions of Eqs.~\eqref{eq:HME1}-\eqref{eq:BC_HME} can be found by setting $E(t,\theta)=E_s(t)e^{-i\omega\theta}$, which adds a contribution $i\omega E$ to the right-hand side of Eq.~\eqref{eq:HME1}. Here, the spectral parameter $\omega$ is a free parameter that is automatically adapted during the continuation. In order to determine $\omega$, we impose an auxiliary condition that breaks the phase-shift symmetry of the system. The dynamical boundary condition Eq.~\eqref{eq:BC_HME} is implemented by adapting the singular mass matrix of the corresponding generalized eigenvalue problem taking into account the contributions from a single boundary element.

\begin{figure}[htbp]
		\centering \includegraphics[width=1.0\columnwidth]{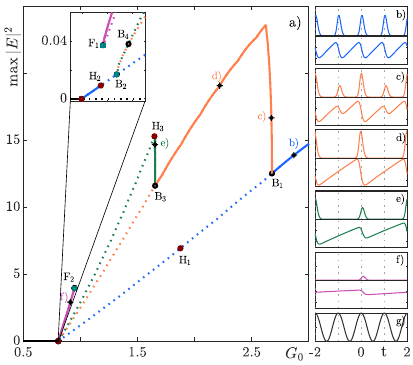}
		\caption{(a) Bifurcation diagram of stationary solutions of the HME given by Eqs.~\eqref{eq:HME1}-\eqref{eq:BC_HME} for $t\in(0,\,4)$ showing the maximum field intensity as a function of $G_0$ obtained for $\alpha=0$ at the fixed value of the detuning $\upsilon=-0.005$. Stability is indicated with thick lines. Open, red and teal circles indicate the positions of the pitchfork, Andronov-Hopf and fold bifurcation points, respectively. Black, blue, orange, green and purple branches correspond to the off, HML$_4$, TC$_2$, symmetry-broken TC$_2$ and TC$_4$ solutions, respectively. The black square marks the close-up around the threshold. (b-f) Exemplary profiles of the fields $E$ (top) and $G$ (bottom) calculated at the labeled positions in (a). (g) Modulation profile with $\omega_m=2\pi$. Vertical dotted lines indicate the positions of the modulation maxima. Other parameters are: $(\gamma,\, k,\,\Gamma,\,m)=(40,\, 0.916,\,0.1,\,0.1)$.}\label{fig:P2P_G0V_m0005}
	\end{figure}

We first consider $\alpha=0$ and fix the detuning to a small blue detuning value of $\upsilon=-0.005$, where the parabolic approximation of the potential holds. The results of the path-continuation using the gain value $G_0$ as a control parameter are presented in Fig.~\ref{fig:P2P_G0V_m0005}~(a), whereas panels (b-f) show exemplary field intensity (top) and gain (bottom) profiles, and panel (g) depicts the corresponding modulation profile.

One can see that for high gain values an HML$_4$ solution with four pulses is stable, cf. Fig.~\ref{fig:P2P_G0V_m0005}~(b). We note that, as expected, the pulses are lagging behind the modulation and that they are slightly shifted to the right of the modulation maxima. Decreasing $G_0$, this solution becomes unstable at the supercritical pitchfork bifurcation point $B_1$. At this point, two stable symmetrical solution branches emerge, where the intensity of pulses at even (odd) maxima starts to decrease with decreasing $G_0$, see Fig.~\ref{fig:P2P_G0V_m0005}~(c). At the same time, the intensity of the remaining odd (even) pulses increases and reaches its maximum when the other pulses disappear. This results in two discrete time-crystal-two (TC$_2$) configurations, where each pulse is followed by an empty slot before the next pulse emerges~\cite{WKY_PRL_26}, cf. Fig.~\ref{fig:P2P_G0V_m0005}~(d). Note that the TC$_2$ states, corresponding to even (odd) pulse patterns, are degenerate and indistinguishable in the measure chosen in Fig.~\ref{fig:P2P_G0V_m0005}~(a) and in what follows we shall study the behavior of only one of those branches. With a further decrease in $G_0$, the formed TC$_2$ solution loses stability at another supercritical pitchfork bifurcation point $B_3$. Here, two stable solution branches of symmetry-broken TC$_2$ emerge, where the intensity of one of the pulses decreases with decreasing $G_0$, see Fig.~\ref{fig:P2P_G0V_m0005}~(e). This solution loses stability at an Andronov-Hopf (AH) bifurcation point $H_3$. The unstable branch then reconnects with the TC$_2$ at the pitchfork bifurcation point $B_4$ (see the inset in Fig.~\ref{fig:P2P_G0V_m0005}~(a)). The latter, in turn, reconnects to the unstable HML$_4$ solution, which stabilizes at an AH point $H_2$ (note also another AH point $H_1$ on the HML$_4$ branch). The intensity of the resulting HML$_4$ solution decreases with decreasing $G_0$ until it reconnects with the off-state at $G_0=G_{\mathrm{th}}$, given by Eq.~\eqref{eq:Gth_0}.

At small gain values, another isolated branch exists (cf. magenta curve in Fig.~\ref{fig:P2P_G0V_m0005}~(a)). The corresponding solutions are stable between two fold points $F_{1,2}$, and consist of one pulse followed by three empty slots, thus forming a time-crystal-four (TC$_4$) solution, cf. Fig.~\ref{fig:P2P_G0V_m0005}~(f).

	\begin{figure}[htbp]
		\centering \includegraphics[width=1\columnwidth]{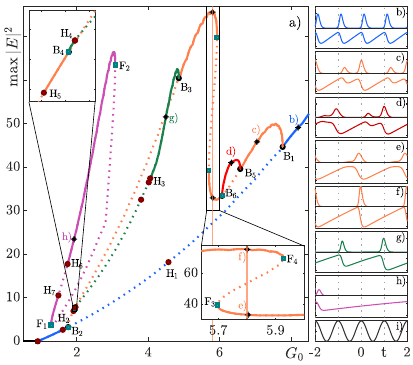}
		\caption{(a) Bifurcation diagram of stationary solutions of the HME given by Eqs.~\eqref{eq:HME1}-\eqref{eq:BC_HME} for $t\in(0,\,4)$ showing the maximum field intensity as a function of $G_0$ obtained for $\alpha=0$ at the fixed value of the detuning $\upsilon=-0.016$. Stability is indicated with thick lines. Open, red and teal circles indicate the positions of the pitchfork, AH and fold bifurcation points, respectively. Black, blue, orange, green and purple branches correspond to the off, HML$_4$, TC$_2$, symmetry-broken TC$_2$ and TC$_4$ solutions, respectively. The black square marks the close-up around the threshold. (b-f) Exemplary profiles of the fields $E$ (solid) and $G$ (dashed) calculated at the labeled positions in (a). (g) Modulation profile with $\omega_m=2\pi$. Vertical dotted lines indicate the positions of the modulation maxima. Other parameters as in Fig.~\ref{fig:P2P_G0V_m0005}.}\label{fig:P2P_G0V_m0016}
	\end{figure}
Next, we performed a continuation for a larger detuning value of $\upsilon=-0.016$. In the resulting bifurcation diagram, shown in Fig.~\ref{fig:P2P_G0V_m0016}, we observe a similar morphology of emerging solutions: a stable HML$_4$ solution (blue, panel (b)) for small and large values of $G_0$ and a pair of pitchfork bifurcations at the points $B_{1}, B_{2}$, leading to the emergence of TC$_2$ states (orange, panel (c)). The latter experience further symmetry-breaking bifurcations (green) at points $B_3$ and $B_4$ (cf. the inset of panel (a) and the exemplary profile in panel (g)).

In contrast to the case of smaller $\upsilon$ values, which is depicted in Fig.~\ref{fig:P2P_G0V_m0005}, the new branch of TC$_2$ states experiences a pair of further symmetry-breaking pitchfork bifurcations at the points $B_5$, $B_6$, resulting in a stable symmetry-broken TC$_2$ solution, where each pulse possesses a small satellite (cf. Fig.~\ref{fig:P2P_G0V_m0016}~(d)). Another difference from the case presented in Fig.~\ref{fig:P2P_G0V_m0005} becomes visible in the inset of Fig.~\ref{fig:P2P_G0V_m0016}~(a): The TC$_2$ solution possesses a small region of bistability between the folds $F_3$ and $F_4$, see Fig.~\ref{fig:P2P_G0V_m0016}~(e,f). Finally, the branch of TC$_4$ solutions (magenta) constitutes an isolated loop as in the case of smaller detuning and experiences an additional region of AH instability between AH points $H_7$ and $H_8$, cf. Fig.~\ref{fig:P2P_G0V_m0016}~(h).

\begin{figure}[htbp]
		\centering \includegraphics[width=1\columnwidth]{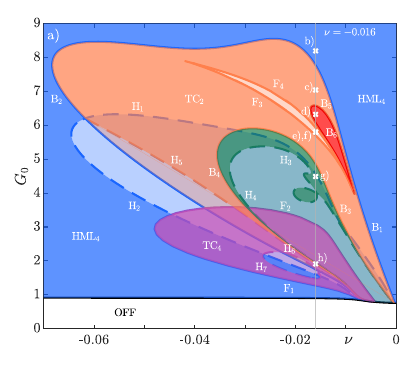}
		\caption{(a) Two-parameter bifurcation diagram in the $(\upsilon,\,G_0)$ plane, showing the evolution of the bifurcation points marked in Fig.~\ref{fig:P2P_G0V_m0005}, Fig.~\ref{fig:P2P_G0V_m0016}. Solid (dashed) lines delimit the existence (stability) regions. Bright (pale) colors indicate stable (unstable) regions. The white region indicates the stable off solution. The labels (b-h) mark the positions of solution profiles, depicted in Fig.~\ref{fig:P2P_G0V_m0016}, cf. Supplementary Video 1. Other parameters as in Fig.~\ref{fig:P2P_G0V_m0005}.}\label{fig:P2P_2param}
	\end{figure}
	\begin{figure}[htbp]
		\centering \includegraphics[width=1\textwidth]{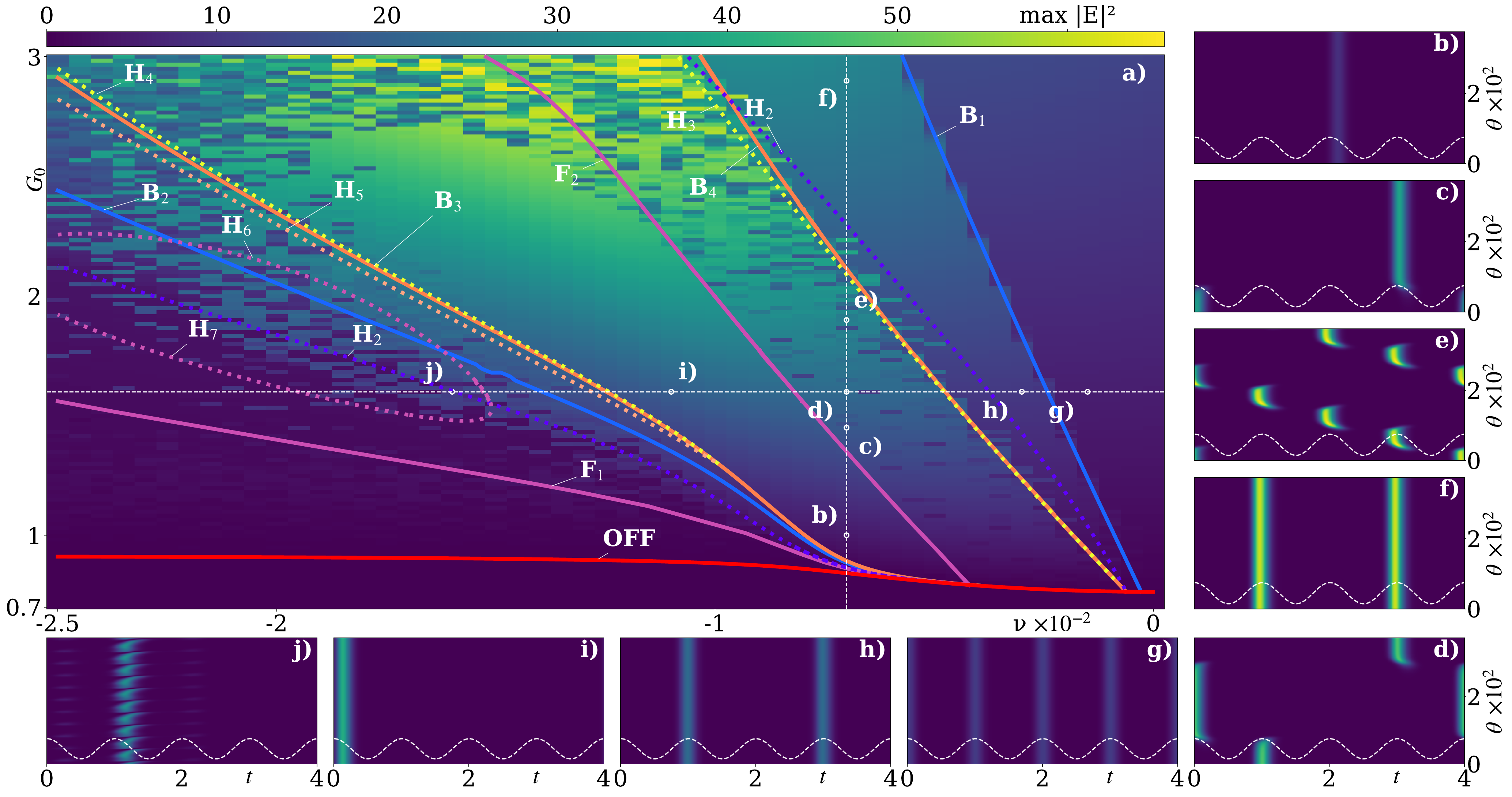}
		\caption{(a) The bifurcation diagram in the $(\upsilon,\,G_0)$ plane obtained with direct numerical simulations of the HME given by Eqs.~\eqref{eq:HME1}-\eqref{eq:BC_HME} for $\alpha=0$. The colormap corresponds to $\max(|E|^2)$. The colored lines correspond to the bifurcation lines presented in Fig.~\ref{fig:P2P_2param}. (b-j) Exemplary field intensities in a two-time representation calculated at points along white lines at $G_0=1.6$ and $\upsilon=-0.007$, respectively. An excerpt of the last 400 round trips of a long-time simulation over $2\times 10^5$ round trips is presented. Other parameters as in Fig.~\ref{fig:P2P_G0V_m0005}.}\label{fig:dns_2param}
	\end{figure}

To better identify the existence and stability regions of the different solution branches, we perform a two-parameter continuation in the $(\upsilon,\,G_0)$ plane of all bifurcation points identified in Figs.~\ref{fig:P2P_G0V_m0005} and ~\ref{fig:P2P_G0V_m0016}. This results in the bifurcation diagram presented in Fig.~\ref{fig:P2P_2param}. Here, solid and dashed lines correspond to the existence and stability of corresponding solutions, colored in the same way as the solution branches in Figs.~\ref{fig:P2P_G0V_m0005} and ~\ref{fig:P2P_G0V_m0016}. The bright (pale) colors correspond to stable (unstable) solution regions. The labeled points (b-h) on the vertical line at $\upsilon=-0.016$ correspond to the position of exemplary solution profiles, depicted in Fig.~\ref{fig:P2P_G0V_m0016}. The evolution of the solution profile along the line is presented in Supplementary Video 1.

The white region below the black threshold line indicates the stable off-state. The threshold itself is parabolic for small $\eta$ as predicted by Eq.~\eqref{eq:Gth_0}, while it approaches a constant line for large $\upsilon$ at $G_{\mathrm{th}}=k$. One can see that the TC$_2$ solution possesses a large existence and stability region, whereas the TC$_4$ states are restricted to smaller gain and detuning values. In addition, the region of TC$_2$ bistability represents a tiny loop at high gain values, limited by the lines $F_3$ and $F_4$. Our results indicate a complex coexistence scenario of the different types of solutions in parameter space, which complicates the numerical and experimental search for these states.

\subsection{Time simulations of the Haus master equation}

While the behavior of different steady-state solutions of the HME~\eqref{eq:HME1}-\eqref{eq:BC_HME} in parameter space can be studied with bifurcation analysis, as presented in Figs.~\ref{fig:P2P_G0V_m0005}-\ref{fig:P2P_2param}, time simulations can provide additional information about a system's temporal behavior while also providing a basis for comparison and confirming the results obtained from path-continuation. To that end, we conducted long-time simulations over $2\times 10^5$ round-trips of the HME~\eqref{eq:HME1} with the asynchronous boundary condition Eq.~\eqref{eq:BC_HME} for the same parameter set as in the previous section. A split-step method with $N_x=256$ grid points on the domain $t\in(0,\,4)$ was employed for the numerical integration. As an initial condition, a weakly white-noise-perturbed off-state was used. The results of a numerical scan in the $(\upsilon,\,G_0)$ plane are presented in Fig.~\ref{fig:dns_2param}. Here, the colormap represents the field intensity maximum. Further, two white dotted lines indicate two cuts through the plane at $G_0=1.6$ and $\upsilon=-0.007$, respectively. The colored lines correspond to the bifurcations presented in Fig.~\ref{fig:P2P_2param}, indicating good agreement between the time simulations and the path-continuation results. Finally, the labeled points along the white lines mark the position of the corresponding $E$-field intensity profiles, shown in a two-time representation in panels (b)-(j). Here, excerpts of the last 400 round-trips are presented. Along the white lines, we observe a stable HML$_4$ solution above the $B_1$ line in panel (g) and stable TC$_2$ solutions in panels (f) and (h) below $B_1$, as expected. Further, a stable TC$_4$ solution exists between the $F_1$ and $F_2$ lines, see panels (b) and (i). The latter can be unstable with respect to an AH bifurcation, cf. panel (j). Finally, different complex periodic regimes, presented in panels (c), (e), and (d) exemplify the dynamics in the region between the lines spanned by the AH and pitchfork bifurcations of HML$_4$ ($H_1$, $H_2$), TC$_2$ ($B_3$, $B_4$) as well as symmetry-broken TC$_2$ ($H_3$, $H_4$), where none of the steady-state solutions are stable (cf. Fig.~\ref{fig:P2P_2param}).

Notice that the bifurcation analysis presented in Figs.~\ref{fig:P2P_G0V_m0005}-~\ref{fig:P2P_2param} and the numerical simulations in Fig.~\ref{fig:dns_2param} were conducted for $\alpha=0$. Figure~\ref{app:fig:fig:dns_2param_alp} in Appendix~\ref{app:bifurcation} presents a numerical bifurcation diagram in the $(\upsilon,\,G_0)$ plane obtained for a more realistic value of $\alpha=1.5$. One can see that the overall morphology of the emerging solutions remains the same as for $\alpha=0$, and one can find regions of stable HML$_4$, TC$_2$ as well as TC$_4$ solutions. However, in the region of large gain, complex oscillatory dynamics set in, reducing the region of stability of, e.g., HML$_4$ and TC$_2$ states.



\section{Effective equations of motion for active mode-locking}

Numerical results of the HME~\eqref{eq:HME1}-\eqref{eq:BC_HME} indicate that the interplay between the modulation, which defines the position of the pulse in the potential, and the gain dynamics plays an important role in the formation of the TC phases. Further, TC states are only observed if the modulation frequency is faster than the natural one, effectively reducing the time for gain recovery between adjacent pulses and forcing pulses to settle on the falling edge of the modulator transmission, which is an unfavorable position from the point of view of the round-trip losses. However, the pulse positions evolve so slowly that resolving their dynamics requires simulations over several tens of thousands of round trips. This motivates the derivation of an effective EOM model to study the asymptotic pulse dynamics. Its formulation as a system of coupled ordinary differential equations also provides additional insight into the mechanisms underlying TC formation.

\begin{figure}[htbp]
	\centering \includegraphics[width=1\columnwidth]{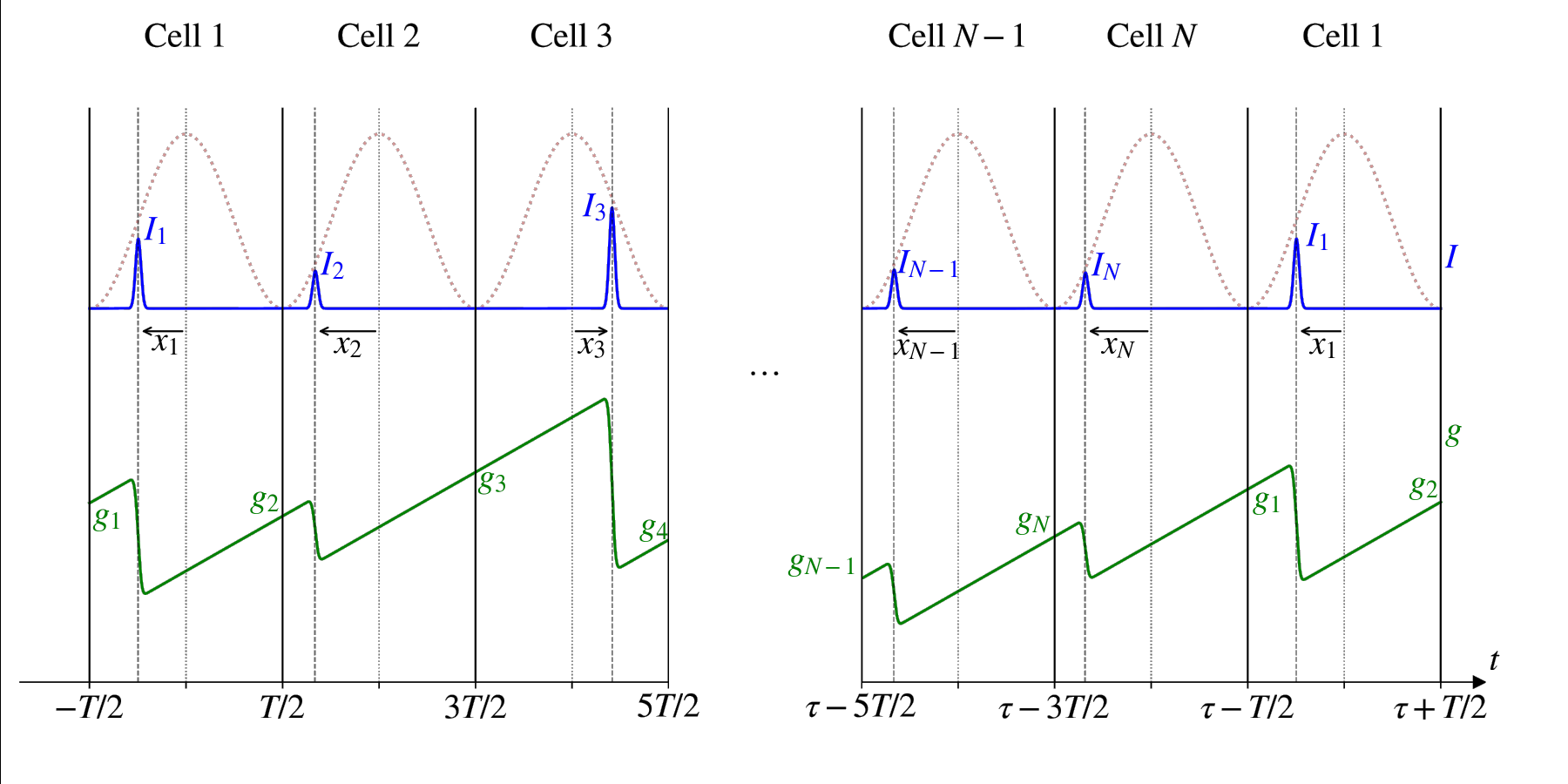}
	\caption{A chain of $N$ well-separated pulses (blue) in a cavity of length $\tau$ occupies $N$ cells of length $T$ created by the modulation (dashed black). The pulses are characterized by their respective intensities $S_j$, displacement $y_j$ (black) with respect to the transmission maximum at $t_j=jT$ and gain variation in each slot $g_j$ (green).}
    \label{fig:Sketched_model}
\end{figure}

To this end, we consider an AML system operating in the HML$_N$ regime for $\upsilon\ll1$. This results in a sequence of $N$ well-separated Gaussian pulses, given by Eq.~\eqref{eq:HGM0}, in a cavity of length $\tau$. Each of the $N$ pulses occupies a consecutive position within the cells of length $T$ imposed by the modulation, as schematically illustrated in Fig.~\ref{fig:Sketched_model}. Each pulse is described by its amplitude $\rho_j$ (with the corresponding integrated intensity $I_j=\rho_j^2$ due to our choice of normalization for $h_0$) and a small displacement $x_j$ with respect to the center of the cell located at the transmission maxima $t_j=jT$. The corresponding gain variations along the cavity due to the presence of the pulses are described with the variables $g_j$.
The pulses' degrees of freedom $(I_j, x_j)$ are functions of the slow time $\theta$ and we employ the ansatz
\begin{equation}
 E(t,\theta)=\sum_{j=0}^{N-1}\sqrt{I_j(\theta)}h_0(t-t_j-x_j(\theta))\,,
\end{equation}
 for the field $E$. As in Ref.~\cite{PGG-NAC-20}, we decompose the gain $G$ into a slowly evolving mean field $\widebar{G}(\theta)=\frac{1}{\tau}\int_0^{\tau}G(t,\theta)\,dt$ and a fast, zero-average contribution $g(t,\theta)$,
\begin{equation}
G(t,\theta)=\widebar{G}(\theta)+g(t,\theta)\,. \label{eq:Gdecomp}
\end{equation}
In our model, which neglects effects present in dispersive nonlinear waveguides such as the Kerr effect, electrostriction, and Cherenkov (dispersive wave) radiation, the pulse-to-pulse temporal variation of the gain $g(t,\theta)$ is the main mechanism mediating interactions between pulses along the cavity axis. We further define a smallness parameter $\varrho=\Gamma T$. The latter is a direct measure of the amplitude of the gain recovery dynamics between pulses. In the following, we shall consider the regime of high-frequency active mode-locking, which means modulation frequencies at least of the order of 10~GHz in semiconductor lasers whose carrier lifetime is in the nanosecond range. We note that our theory can also be applied to slow gain media such as erbium whose carrier lifetime lies within the 10~ms range. In that case, the high-frequency regime is easily reached at 1~kHz. Assuming that $\varrho \ll 1$ allows us to obtain analytical expressions for $\widebar{G}(\theta)$ and $g(t,\theta)$ that is correct up to $\mathcal{O}(\varrho^2)$, see Appendix~\ref{app:derivation} for more details. In addition, one can easily observe that $g \sim \mathcal{O}(\varrho)$.

Using the Gaussian superposition ansatz coupled to the induced gain dynamics and the element-wise orthogonality of a well-separated superposition of Gaussian functions and of their derivatives, projection yields a system of $2N$ ordinary differential equations for the position and integrated intensity of each pulse, coupled to a recursive relation linking the gain variation in the $j$-th slot to the previous one, thereby preserving the structure of Eq.~\eqref{eq:HME2}. The resulting EOM model, which is derived in more detail in Appendix~\ref{app:derivation}, reads, after rescaling,
\begin{align}
	\dot{S}_j=&\bigg[
	\widetilde{G}-1
	-2\pi\mu r^2 y_j^2+\varrho\left(\hat{g}_j+\widetilde{G}\left[\braket{S}\left(y_j+\frac{1}{2}\right)-\frac{S_j}{2}\right]\right)\bigg]S_j \,, \label{EOM1}\\
	\frac{1}{\mu}\dot{y}_j =&-\tilde{\upsilon}-y_j+\frac{\varrho}{4\pi\mu r^2}\widetilde{G}\left(\braket{S}-r\,S_j\right)\,, \label{EOM2}\\
	\hat{g}_1=&\,\widetilde{G}\sum_{j=1}^N S_j\frac{N+1-2j-2y_j}{2N}\,, \label{EOM3}\\
	\hat{g}_{j+1}=&\,\hat{g}_j+\widetilde{G}(\braket{S}-S_j)\,.
    \label{EOM4}
\end{align}
In Eqs.~\eqref{EOM1}-\eqref{EOM4}, we define the pulse timing relative to the modulator period as $y_j=x_j/T$ and the scaled intensity $S_j=I_j/T$. In addition, the minimal eigenvalue of the Sturm-Liouville problem $\lambda_0^0=\sqrt{m}\omega_m/\gamma$, obtained at resonance when $\upsilon=0$, leads to the minimal round-trip losses and threshold value that we define as $l=k-2m+\lambda_0^0$. This value is used to normalize the local gain in the $j$-th cell as $\hat{g}_j=g_j/(lT)$ and the pumping with respect to the threshold as $J=G_0/l$ while the dot represents the derivative with respect to the normalized time $\theta l$.
Up to first order in $\varrho$, the scaled average gain is $\widetilde{G}=\frac{J}{1+\langle S \rangle}$, where $\langle S \rangle=\frac{1}{N}\sum_j S_j$ thereby confirming that the average gain depends only on the mean field intensity. We note that this expression represents a mean field coupling between pulses. If this were the only nonlinearity present in Eqs.~\eqref{EOM1}-\eqref{EOM4}, the resulting dynamical system would be highly degenerate and potentially unstable since antiphase pulse fluctuations leaving $\braket{S}$ constant would be undamped (neutral) modes. We define $\tilde{\upsilon}=\upsilon/(\lambda_0^0 T)$ as the scaled detuning value normalized to the period $T$ and the restoring force at the modulation transmission maxima $\lambda_0^0$. The parameter $\mu=\lambda_0^0/l$ defines the position relaxation speed normalized to the round-trip cavity losses and, finally, $r=T\sqrt{\beta/\pi}=\sqrt{\gamma T} m^{1/4}$ represents the ratio of modulation period to pulse width, cf. Appendix~\ref{app:derivation}. Pulsed operation corresponds in general to large values of $r$.

In the evolution of the $j$-th pulse intensity given by Eq.~\eqref{EOM1}, the linear term $(\widetilde{G}-1)$ describes the competition between the mean round-trip gain and normalized cavity losses, while the contribution $-2\pi\mu r^2$ accounts, in the parabolic approximation, for the additional losses imposed by the pulses' displacement away from the center of the slot and from the maxima of transmission of the modulator. Note that this displacement does not impose an asymmetric profile, due to the assumption of a Gaussian pulse shape, limiting the validity of the EOM model to small detunings where the parabolic approximation holds. The remaining terms describe the interaction through the gain dynamics.

Further, Eq.~\eqref{EOM2} determines the evolution of the $j$-th pulse position. Here, we identify a restoring force towards the transmission maximum located at $y_j=0$, together with the detuning $\tilde{\upsilon}$ playing the role of a constant force. The last term is essential to the time-crystal formation dynamics as it describes the pulse displacement caused by gain depletion. As the leading edge of a pulse experiences a higher gain than the trailing edge, gain depletion effectively results in a pulse acceleration and therefore a drift to the left in the reference frame of the modulator. This drift increases with pulse intensity. 

If the modulation frequency is blue-detuned, i.e., faster than the natural frequency of the resonator ($\tilde{\upsilon}<0$), the pulses lag and move to the right of the transmission maximum since $y_j=-\tilde{\upsilon}$ in the absence of nonlinear contributions in Eq.~\eqref{EOM2}. The drift induced by the gain can counteract this lag to a certain extent since $r\gg1$, so that the last contribution on the right-hand side of Eq.~\eqref{EOM2} is negative.

Alternatively, if the modulation gets too fast, every other pulse can be eliminated. As a consequence, the remaining pulses become approximately twice as intense, which similarly increases the counteracting effect of the nonlinear drift. This mechanism provides an intuitive understanding of how the surviving pulses may move closer to the transmission maximum. This is the main mechanism underlying time-crystal formation. It also explains why TCs only form on the blue side of the mode-locking resonance tongue, since the drift induced by the gain profile can only accelerate the pulse and thus only counteract a situation where the modulation is too fast.

Finally, Eqs.~\eqref{EOM3}-\eqref{EOM4} determine the gain variations associated with the individual pulses. They describe the combined gain depletion and recovery at each potential slot. In particular, if one pulse, say $S_j$, is smaller than the mean pulse intensity $\langle S\rangle$, the corresponding gain for its right neighbor $\hat{g}_{j+1}$ will increase relative to that of the preceding pulse. Similarly, if the pulse intensity is larger than the average value, the gain depletion exceeds the recovery, reducing $\hat{g}_{j+1}$. In this way, the gain variables mediate an effective interaction between all pulses in the cavity, usually resulting in homogenization of pulse intensities. Note that the value of the gain $\hat{g}_{1}$ in the first slot given in Eq.~\eqref{EOM3} is obtained by imposing periodic boundary conditions.

\subsection{Bifurcation analysis of the effective equations of motion}

\begin{figure}[htbp]
	\centering \includegraphics[width=1\columnwidth]{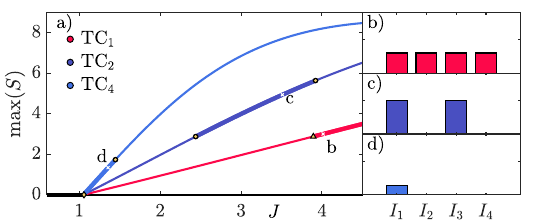}
	\caption{A bifurcation diagram obtained from the EOM~\eqref{EOM1}-\eqref{EOM4}, showing the maximum intensity as a function of $J$. The different stationary solutions HML (red), TC$_2$ (blue) and TC$_4$ (light blue) bifurcate from the off state (black) at a degenerate bifurcation point (diamond). Stable (unstable) solutions are marked with thick (thin) lines. (b-d) show the $S_j$ components of exemplary profiles from the stable regions of the different branches. Parameters are $(\rho,\,\tilde{\upsilon},\,\mu,\,r)=(0.1,\,-0.1,\,0.0646,\,3.5637)$.}
    \label{fig:EOM_TC_bifurcation_reduced}
\end{figure}

At the lasing threshold, the off-state $(0,\,-\tilde{\upsilon},\,0)$ of the EOM model~\eqref{EOM1}-\eqref{EOM4} becomes linearly unstable and $\widetilde{G}=J$, yielding a condition for the scaled pumping rate, $J_\mathrm{th}=1+2\pi\mu\,r^2\,\tilde{\upsilon}^2\,,$ which, after renormalization, coincides with the value $G_0=G_{\mathrm{th}}$ of the HME~\eqref{eq:HME1} for $\upsilon\ll1$, cf. Eq.~\eqref{eq:Gth_0}.
We note that HML and TC states are steady-state solutions of the EOM~\eqref{EOM1}-\eqref{EOM4}. Similar to the analysis of the HME~\eqref{eq:HME1}-\eqref{eq:HME2}, we consider a modulation admitting four slots and continue the steady states of Eqs.~\eqref{EOM1}-\eqref{EOM4} in parameter space, using the numerical continuation framework \textsc{DDE-Biftool}~\cite{DDEBT}. Note that one can find the HML and TC branches by constructing initial guesses based on the symmetry of the states. In particular, HML or TC$_2$ solutions fulfill $S_1=S_2=S_3=S_4=S\neq0$ or $S_1=S_3=S\neq0$ and $S_2=S_4=0$, respectively. The remaining variables are constructed from these assumptions. The steady states of all regular TC configurations can be obtained analytically. The resulting bifurcation diagram is presented in Fig.~\ref{fig:EOM_TC_bifurcation_reduced}~(a), where $\mathrm{max}(S)$ is displayed as a function of $J$.

Further, three exemplary profiles at positions marked by white crosses are shown in panels (b-d), where each pulse is described only by its peak amplitude, reducing the profile to four amplitude values. Remarkably, all three branches emerge from the same degenerate bifurcation point (diamond) on
the off-state branch (black). This is similar to the emergence scenario observed for the HME, although in that case it remained unclear whether the branches truly originated from the same point or merely approached it closely in the case of vanishing detuning.

Just above threshold, only the TC$_4$ phase (light blue) is stable. For larger $J$, one observes a parameter region in which none of the TC states is stable, before entering a regime where the TC$_2$ solution (blue) gains stability. Finally, the HML$_4$ branch (red), hereafter denoted TC$_1$, becomes stable, while the TC$_2$ state loses stability again. This reproduces the hierarchy and fixed order of the steady states, arranged like a staircase, which was already observed for the HME. The points limiting the stability of the TC$_2$ and TC$_4$ solutions (circles) correspond to a bifurcation where a real eigenvalue passes through zero, indicating either a pitchfork or a transcritical bifurcation. The HML branch becomes stable at a bifurcation point (triangle), where a complex-conjugate pair of eigenvalues crosses the imaginary axis simultaneously with an additional real eigenvalue.

\begin{figure}[htbp]
	\centering \includegraphics[width=1\columnwidth]{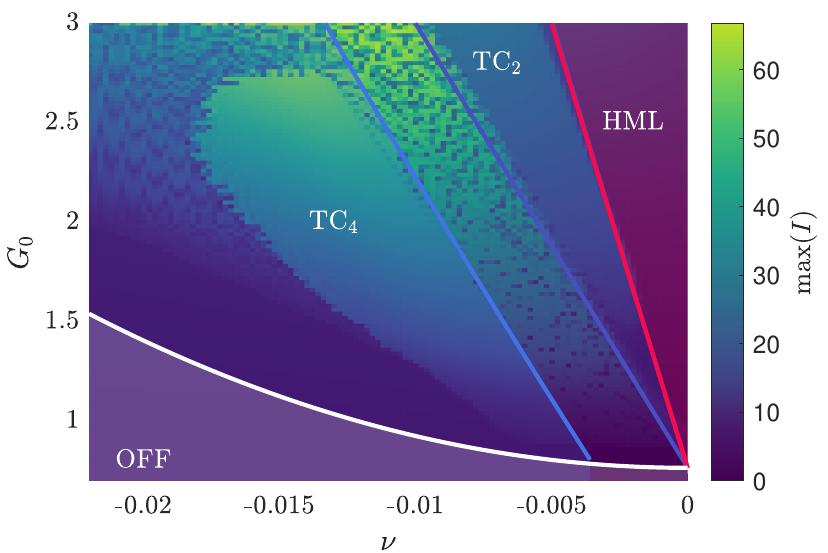}
	\caption{A two-parameter bifurcation diagram obtained from the EOM
	~\eqref{EOM1}-\eqref{EOM4} superimposed on a numerical scan of the HME~\eqref{eq:HME1} for the maximum field intensity $I=|E|^2$ on the $(\upsilon,G_0)$ plane. Red, dark blue and light blue lines correspond to the stability boundaries of HML (TC$_1$), TC$_2$ and $TC_4$ regions, respectively, while the white line corresponds to the threshold. Parameters as in Fig.~\ref{fig:EOM_TC_bifurcation_reduced}.}
    \label{fig:G_0_nu_twopar_diagram}
\end{figure}

To compare the results of the EOM~\eqref{EOM1}-\eqref{EOM4} with those obtained for the HME~\eqref{eq:HME1}, we conducted a two-parameter continuation in the $(\upsilon,\,G_0)$ plane of the bifurcation points limiting the stability of all stationary solutions presented in Fig.~\ref{fig:EOM_TC_bifurcation_reduced}, using the same parameter scaling as in the HME~\eqref{eq:HME1}. The results are presented in Fig.~\ref{fig:G_0_nu_twopar_diagram} where the red, dark blue and light blue lines represent the stability boundaries of HML$_4$/TC$_1$, TC$_2$ and $TC_4$ regions, respectively, while the white line represents the threshold line. The colormap corresponds to the maximum field intensity $I=|E|^2$ obtained with numerical integration of the HME~\eqref{eq:HME1} for the same parameters as in Fig.~\ref{fig:dns_2param}. In Fig.~\ref{fig:G_0_nu_twopar_diagram}, a close-up of the parabolic region at small values of $\upsilon$ is shown. The results are in perfect agreement for small detuning values and close to the threshold, where the EOM system~\eqref{EOM1}-\eqref{EOM4} was derived. One can see that beyond the existence of HML and TC phases, the numerical results for the HME~\eqref{eq:HME1} predict regimes of time-periodic solutions. To confirm their existence in the EOM~\eqref{EOM1}-\eqref{EOM4}, we continue our study by investigating the secondary branches arising from the bifurcation points in Fig.~\ref{fig:EOM_TC_bifurcation_reduced}.

\begin{figure}[htbp]
	\centering \includegraphics[width=1\columnwidth]{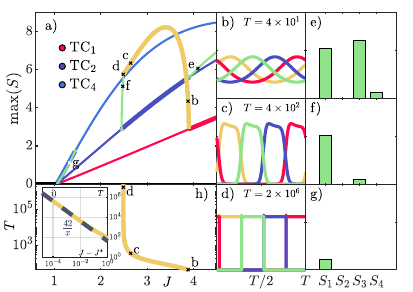}
	\caption{(a) Bifurcation diagram with maximum intensity as a function of $J$. Parameters as in Fig.~\ref{fig:EOM_TC_bifurcation_reduced}. The new branches are the green and black unstable steady state branches, as well as the branch of periodic solutions (yellow), which is stable. (b-d) Profiles from different positions along the branch of periodic solutions. One can see how the period increases as (d) is approached. (e-f) Exemplary profiles from the unstable steady state branches (green). (h) The period $T$ of the branch of periodic solutions as a function of $J$. The inset (i) shows the respective scaling to be $\frac{42}{x}$, hinting at a global bifurcation.}
    \label{fig:EOM_TC_bifurcation}
\end{figure}

This results in the extended bifurcation diagram of the EOM~\eqref{EOM1}-\eqref{EOM4}, displayed in Fig.~\ref{fig:EOM_TC_bifurcation}~(a), where the current $J$ is again the control parameter. Here, green branches correspond to unstable steady states, emerging from the bifurcation points limiting the stability of TC$_4$ and TC$_2$ solutions (cf. circle markers in Fig.~\ref{fig:EOM_TC_bifurcation_reduced}). Further, the stable yellow branch is a branch of periodic solutions, emerging from the HML (TC$_1$) (cf. triangle marker in Fig.~\ref{fig:EOM_TC_bifurcation_reduced}). In Fig.~\ref{fig:EOM_TC_bifurcation}~(b-d), exemplary profiles along the periodic branch at the positions marked with crosses in panel (a) are depicted. Panel (b) shows a periodic orbit shortly after the bifurcation point, indicating the characteristic birth of a sinusoidal, small-amplitude orbit with a period of $T\approx 40$. Even though points (c) and (d) are placed close together, the period increases from $T\approx 400$ up to $T\approx 2\times 10^6$ between the points and we can see that periodic orbits consist of plateau-like square pulses, separated by sharp transition layers, each accompanied by a small overshoot.

Panel (h) shows the corresponding oscillation period $T$ as a function of $J$, indicating a global bifurcation. The scaling of the period as a function of the distance to the global bifurcation point, shown in the inset of Fig.~\ref{fig:EOM_TC_bifurcation}~(h), is $\sim 1/x$. This fit does not correspond to the classical scaling of either homoclinic or saddle-node infinite-period (SNIPER) bifurcations. Notably, in Ref.~\cite{CSA_ScRep_18} an infinite-period scenario involving a transcritical instead of a saddle-node bifurcation has been reported, which led to analogous scaling behavior. Since an unstable
steady-state branch (cf. green line in panel (a), along which point (f) is found) connects to the TC$_4$ branch via a transcritical bifurcation close to the global bifurcation point where the yellow branch terminates, this scenario appears plausible.

Note that the unstable steady-state branch (green, along which point (g) is found) in the lower-left region of Fig.~\ref{fig:EOM_TC_bifurcation}~(a) emerges subcritically at the point where the TC$_4$ branch loses stability, returning to the degenerate bifurcation point at the off-state.
Upon emergence, the pulse amplitude slightly decreases, leading to intensity being distributed into a neighboring slot, as shown in panel (g). A similar mechanism
is observed in panel (f) for the green branch near its bifurcation with the TC$_4$ state, where intensity is transferred to the next-nearest slot. Along this branch towards its connection with the TC$_2$ state, the small pulse grows at the expense of the larger one until both amplitudes become equal at the bifurcation point, corresponding to the TC$_2$ profile. The second green branch originates from the TC$_2$ state and a corresponding profile is shown in panel (e), where a third, initially small pulse emerges.

\subsection{Winnerless competition}

\begin{figure}[htbp]
	\centering \includegraphics[width=1\columnwidth]{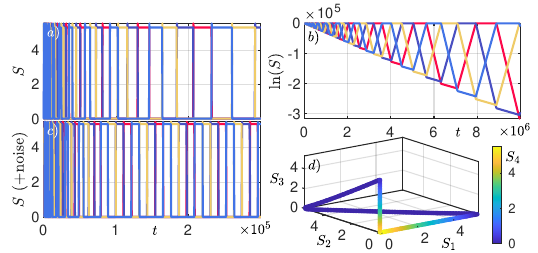}
	\caption{(a) Numerical simulation of EOM~\eqref{EOM1}-\eqref{EOM4} in the WLC regime of length $\Delta t=3\times 10^5$. In the absence of noise, the period continues to increase as the trajectory approaches a heteroclinic cycle. (b) The same time trace, but now with a length of $\Delta t=10^7$ on a logarithmic scale. It becomes visible how close the zero equilibrium is approached as the plateaus grow. (c) In the presence of weak noise, the trajectory is pushed away from zero, preventing the infinite growth and ultimately arresting the period. (d) A representation of the trajectory in the $\{S_i\}$ phase space. The fourth variable is indicated by color.}\label{fig:WLC}
\end{figure}

At this point, we can explain most of the branches and connections shown in Fig.~\ref{fig:EOM_TC_bifurcation}. The major remaining question concerns the nature of the region around $J=2$, where none of the branches shown is stable. We note that all bifurcations are connected with their respective branches. Hence, the behavior in this region should either be determined by an isolated branch or connected in some other way. This issue can be addressed by analyzing the time evolution of the EOM~\eqref{EOM1}-\eqref{EOM4} in the area of interest. A long-time simulation over a duration of $\Delta t=3\times 10^5$ started in that area yields the time trace presented in Fig.~\ref{fig:WLC}~(a). There, in the absence of noise, the period continues to increase indefinitely without showing signs of saturation.
Figure~\ref{fig:WLC}~(b) shows the same trajectory over a much longer integration interval of $\Delta t=10^7$ using a logarithmic scale to illustrate how close the plateaus between transitions approach the equilibrium at zero.

The simulation depicted in Fig.~\ref{fig:WLC}~(c) illustrates how the addition of Gaussian white noise prevents the divergence of the period at long times, so that it eventually settles to a finite value. Finally, in Fig.~\ref{fig:WLC}~(d) the evolution in the $\{S_j\}$ phase space is shown. To better visualize the resulting four-dimensional phase portrait, the fourth dimension is encoded by the color scale. We observe a cycle between four states, each characterized by one finite component while the remaining three components are close to zero.
A very slow limit cycle could provide one possible explanation. However, given the enormous timescales involved and the absence of any visible approach to a periodic attractor, we propose a different interpretation of the observed dynamics. Specifically, we argue that the phenomenon can be understood within the framework of heteroclinic networks and the winnerless competition principle (WLC)~\cite{HR_JMAA_09,KM_ETDS_95, K_JNS_97}. Here, the WLC is defined as a stable heteroclinic contour connecting single-species equilibria in a rock-paper-scissors competition, where each species beats exactly one other while losing to another such that the species take turns endlessly without a winner. Meanwhile, the limiting set is not a periodic orbit, but a heteroclinic cycle with no finite period.

Applying this idea to the EOM dynamics at hand, we conclude that the TC$_4$ state, although represented by a single branch, comprises four symmetry-related equilibria. When the equilibrium is unstable, it can be interpreted as a network of four symmetry-related saddle equilibria. As the period diverges at the global bifurcation, the transitions approach heteroclinic connections between these symmetry-related saddle equilibria.

Finally, we note that the divergence of the switching period relies on trajectories approaching
the single-pulse saddle states ever more closely, while the intensities in the remaining slots 
become vanishingly small. In practice, this approach is limited by mechanisms absent from the 
ideal deterministic reduction. As shown above, noise continually seeds the suppressed slots,
preventing arbitrarily close approaches to the invariant subspaces of zero pulse intensity 
and limiting the residence times near the saddles. However, even in the absence of noise, 
finite pulse overlap may provide an additional limitation. Indeed, Gaussian pulses have 
nonzero tails that extend into neighboring modulation slots, so that a nominally single-pulse
configuration does not imply a strictly vanishing field elsewhere in the cavity. These weak 
contributions, neglected in the well-separated-pulse approximation underlying the EOM, 
may seed the growth of suppressed pulses and thereby regularize the ideal heteroclinic dynamics. 
The infinite-period behavior should therefore be understood as a limiting prediction of a 
reduced model, whose realization in the full field dynamics can be limited 
by both noise and residual pulse overlap.

\section{Conclusion}

We have analyzed the multipulse dynamics of an AML semiconductor laser and clarified the physical origin of the subharmonic time-crystalline pulse trains recently observed experimentally in a unidirectional AML ring laser. To this end, we employed an HME generalized to retain the carrier memory from one round-trip to the next, which is the essential ingredient for semiconductor devices whose gain recovery time is comparable to, or longer than, the cavity round trip.

The bifurcation analysis revealed a well-defined hierarchy of steady operating states, organized in a fixed order as the pump current and the modulation detuning were varied: the conventional HML$_N$ state, a period-doubled TC$_2$ phase in which every second slot is unoccupied, and a period-quadrupled TC$_4$ phase, where only one of the four slots is populated. Finally, weakly asymmetric variants of the latter were observed, carrying small satellite pulses. The staircase-like arrangement of these states, their coexistence over sizable regions of the $(\upsilon, G_0)$ plane, and the intervening windows of complex, non-stationary pulse trains were confirmed by long-term numerical integration of the model, and persist for a realistic value of the linewidth enhancement factor.

The mechanism behind this hierarchy is made transparent by the derived effective EOM model, which reduces the field dynamics to a set of coupled ordinary differential equations for the energy of each pulse, its position relative to the corresponding transmission maximum, and the local gain in its slot. This reduction isolates the essential physical competition: the modulator provides a restoring force pulling each pulse toward the transmission maximum, while gain depletion - the leading edge of a pulse seeing more gain than its trailing edge - pushes it forward. When the modulation runs faster than the cavity's natural repetition rate, the pulses lag behind the transmission maxima and pay an additional cost in losses. Suppressing every second (or every fourth) pulse allows the remaining pulses to grow, which increases their gain-induced drift and re-centers them in the modulation window, thereby reducing the overall losses. Subharmonic operation is therefore not an accident of a particular device, but the energetically favorable response of a fast-gain AML laser driven above its natural repetition rate.

The reduced model also accounts for the dynamical regimes that separate the locked states. The boundaries of the TC states correspond to transfer of energy into a neighboring or next-nearest slot, and the transition out of the TC$_4$ regime proceeds through a global infinite-period bifurcation, associated with a transcritical instability. 
Beyond this global bifurcation, trajectories do not settle into a fixed pulse pattern. Instead, the four symmetry-equivalent TC$_4$ configurations become saddle equilibria linked by a heteroclinic network, and the output cycles indefinitely between them in a WLC, with successive plateaus growing ever longer until noise arrests the growth and fixes a finite, noise-determined switching time. This connection to a mechanism familiar from population dynamics can further support the understanding of the complex dynamical regimes that separate the steady-state TC phases, observed in the HME.

From an applications standpoint, discrete time-crystalline operation offers a route to divide the repetition rate of an AML laser by an integer factor without driving the modulator more slowly, thereby decoupling the pulse duration from the pulse spacing - a coupling that has traditionally limited low-repetition-rate active mode locking, where shallow effective potentials yield broad pulses. The two-parameter maps delineate where the different subharmonic phases are stable, where they coexist, and where the device exhibits complex oscillatory dynamics, thus constituting a practical design guide for pulse-picking-free repetition-rate division.


\section*{Funding}
J.J. and S.G. acknowledge the financial support of the project PULPO-AML funded by the Comunitat Autònoma de les Illes Balears through an Ajut per a projectes de recerca científica i tecnològica (PRD2025\_047).

\section*{Declaration of competing interest}
The authors declare no conflicts of interest.

\section*{Data availability}
Data underlying the results presented in this paper are not publicly available at this time but may be obtained from the authors upon reasonable request.

\section*{Supplementary video}
Supplementary Video 1 illustrates the evolution of the solution profile along the line at $\upsilon=-0.016$ in Fig.~\ref{fig:P2P_2param}. The accompanying file is \texttt{video2p.mp4}.

\clearpage
\appendix
\numberwithin{equation}{section}
\numberwithin{figure}{section}
\numberwithin{table}{section}
\section{Derivation of the effective equations of motion}
\label{app:derivation}

Here we derive the effective equations of motion used in the main text to describe the emergence of TCs. We consider $N$ well-separated pulses circulating in the cavity and characterize each pulse by its amplitude and position, together with the gain variation associated with its slot. The reduction yields $2N$ coupled ordinary differential equations (ODEs) for the pulse amplitudes and positions, supplemented by recursive relations for the gain.
\subsection{Gain dynamics}
We begin with the gain dynamics driven by an intensity distribution $I(\theta,t)$:
\begin{align}
	\partial_t G &= \Gamma \left( G_0-G(1+I)\right).\label{app:eq:Gain1}
\end{align}
Motivated by the slow dynamics observed in the TC regimes, we use the steady-state form of the dynamical boundary condition, Eq.~\eqref{eq:BC_HME}. This imposes a periodic boundary condition (PBC) on the gain, $G(\theta,0) = G(\theta,\tau)$. This approximation excludes some instabilities on the slow $\theta$ scale, including the loss of damping of relaxation oscillations, also referred to as self-Q-switching. 
As in Ref.~\cite{PGG-NAC-20}, we decompose the gain into its cavity average and a fluctuation with zero mean:
\begin{align}
	G(\theta,t)=\bar{G}(\theta)+g(\theta,t).
\end{align}
The average $\bar{G}(\theta)$ evolves on the slow round-trip timescale, whereas $g$ describes the fast variations along the cavity axis that mediate pulse interactions. To obtain an explicit expression for these variations, we expand in powers of the gain recovery rate $\Gamma$. 
For an arbitrary quantity $X$, we define its average over one round trip as
\begin{align}
	\bar{X}(\theta)=\braket{X}=\frac{1}{\tau}\int_0^{\tau}X(\theta,t)dt.
\end{align}
The decomposition is consistent with this definition because $g$ has zero average on the fast timescale. Since $\bar{G}$ is independent of the fast time, the PBC for $G$ also applies to $g$. Its evolution equation is
\begin{align}
	\partial_t g&=\Gamma\left[G_0-\bar{G}(1+I)-g-gI\right].\label{app:equation_for_g}
\end{align}
Integrating over the cavity length and using the PBC yields
\begin{align}
	0&=G_0-\bar{G}(1+\bar{I})-\braket{gI},
\end{align}
where $\bar{g}=0$ by definition. Rearranging gives a relation between the average gain and its fluctuation:
\begin{align}
	\bar{G}=\frac{G_0-\braket{gI}}{1+\bar{I}}
\end{align}
Substituting this result into Eq.~\eqref{app:equation_for_g} gives
\begin{align}
	\partial_tg&=\Gamma\left[\bar{G}(\bar{I}-I)+\braket{gI}-g-gI\right].
\end{align}
We now expand $(\bar{G},g)$ perturbatively. The relevant small quantity is the gain recovery over one modulation cell, rather than the dimensional recovery rate alone. Normalizing the fast time to the modulation period $T$ gives the dimensionless parameter $\varrho=\Gamma T$, as in the main text. This parameter is small for erbium lasers in the regime considered here and for semiconductor lasers driven at multi-GHz frequencies. To avoid introducing another time variable, we retain the original notation and write the expansion in powers of $\Gamma \ll 1$, with the smallness condition understood relative to the modulation timescale. We set
\begin{align}
	\bar{G}(\theta)&=\bar{G}_0(\theta)+\Gamma \bar{G}_1(\theta)+\mathcal{O}(\Gamma^2)\\
	g(\theta,t)&=g_0(\theta,t)+\Gamma g_1(\theta,t)+\mathcal{O}(\Gamma^2).
\end{align}
At lowest order, $\partial_t g_0=0$, so $g_0(\theta,t)=g_0(\theta)$. Because the zero-mean condition must hold at each order, $g_0(\theta)=0$. For the average gain, we obtain $\bar{G}_0=G_0/(1+\bar{I})$. The equation for the first nonzero fluctuation is $\partial_t g_2=\bar{G_0}(\bar{I}-I)$. Integrating over the fast time gives
\begin{align}
	g_2(\theta,t)&=\bar{G}_0\left(\bar{I}t-P(t)-c\right).
\end{align}
Here, the cumulative intensity is defined by
\begin{align}
	P(t)&=\int_0^t I(\theta,s)ds.
\end{align}
The identity $P(\tau)=\bar{I}\tau$ ensures that the fluctuation satisfies the PBC. The integration constant $c$ is fixed by requiring the correction to have zero average. Integrating over one round trip gives
\begin{align}
	0&=\bar{G}_0\left(\bar{I}\frac{\tau^2}{2}-\bar{P}\tau-c\tau\right)
\end{align}
leading to $ c = \bar{I}\tau/2-\bar{P}$. Using this result, we can write the solution as
\begin{align}
	g_2(\theta,t)&=\bar{G}_0\left(\bar{I}(t-\frac{\tau}{2})-P(t)+\bar{P}\right). \label{app:g2_equation}
\end{align}
This expression has zero average by construction. For a constant intensity $I=\bar{I}$, it gives $g_2(\theta,t)=0$, as expected. 
Combining the zeroth- and first-order contributions gives the gain profile for $t\in [0,\tau]$:
\begin{align}
	G(\theta,t)=\frac{G_0}{1+\bar{I}}\left[1+\Gamma\left(\bar{I}[t-\frac{\tau}{2}]-P(t)+\bar{P}\right)\right]+\mathcal{O}(\Gamma^2)
\end{align}

For the EOM derivation, it is convenient to express the gain variation between two points along the fast cavity coordinate:
\begin{align}
	g(\theta,t_2)&=g(\theta,t_1)+\bar{G}_0\left(\bar{I}(t_2-t_1)-\int_{t_1}^{t_2} I(\theta,s)ds\right).
\end{align}

\subsection{Haus equation}
\label{app:apx:HME}
\subsubsection{Sturm-Liouville eigenvalue problem}

We start from the Haus model~\cite{H_IEE_75}:
\begin{align}
	\partial_\theta E=\left(\frac{G-k}{2}+m+\upsilon\partial_t+\frac{1}{2\gamma^2}\partial_t^2+m[\cos(\omega_m t)-1]\right)E\label{app:E-equation}
\end{align}

The transmission maxima are located at $t_n=nT$, with $T\omega_m=2\pi$. For narrow pulses, $\gamma\gg 1$, and small detuning, $\upsilon\ll 1$, we expand the cosine around one maximum. The resulting Sturm-Liouville problem determines the mode profiles for a product ansatz:
\begin{align}
	\left(\frac{1}{2\gamma^2}\partial_t^2-\frac{m\omega_m^2}{2}t^2 + L +\upsilon\partial_t\right)h_j(t)=-\frac{\lambda_j}{2}h_j(t) \label{app:eq:SLP}
\end{align}
Here, the $h_j$ are Hermite functions, and $L=\frac{G-k}{2}+m$ is assumed to be small near the balance between gain and losses. Although a displacement of the Gaussian can accommodate drift, sufficiently large drift may deform the mode and invalidate the approximation. The assumption $L\rightarrow 0$ is most appropriate near threshold, where $G$ is nearly constant and close to $k$, provided that the modulation is weak. The lowest-order mode is a Gaussian:
\begin{align}
	h_0(t)=\left(\frac{2\beta}{\pi}\right)^\frac{1}{4}e^{-\beta (t-a)^2}
\end{align}
We normalize the mode to unit integrated intensity:
\begin{align}
	\int_{-\infty}^\infty h_0^2(t)dt=1
\end{align}
Substituting the Gaussian into the eigenvalue problem gives a condition for $(a(\upsilon),\beta,\lambda_0(\upsilon))$:
\begin{align}
	\frac{-2\beta+4\beta^2 (t-a)^2}{2\gamma^2}-2\upsilon\beta(t-a)-\frac{m\omega_m^2}{2}t^2+\frac{\lambda_0}{2}=0.
\end{align}
From this equation, we obtain
\begin{align}
	\beta&=\frac{\sqrt{m}\gamma\omega_m}{2}\\
	\lambda_0(\upsilon)&=\frac{\sqrt{m}\omega_m}{\gamma}+(\gamma\upsilon)^2\\
	a(\upsilon)&=\frac{-\gamma\upsilon}{\sqrt{m}\omega_m}.
\end{align}
The eigenvalue $\lambda_0(\upsilon)$ represents an additional loss: a broader filter reduces this loss, whereas faster modulation increases it. As the pulse width $\tau_p\propto 1/\sqrt{\beta}$ is inversely proportional to $\lambda_0(\upsilon=0)$, we also find that broader pulses $\beta\rightarrow 0$ incur lower losses as $\lambda_0(0)\rightarrow 0$.
The assumption of narrow pulses with a small detuning $\upsilon\ll 1$ leads to a simplified relation for the lasing threshold. At threshold, $G=G_0=\bar{G}$ and $I=0$, giving
\begin{align}
	G_\text{th}=k-2m+\frac{\sqrt{m}\omega_m}{\gamma}+(\gamma\upsilon)^2,
\end{align}
The threshold is therefore parabolic in $\upsilon$. For large detuning, $\upsilon \gg 1$, the drift term cannot be eliminated, and neither the local cosine expansion nor the Gaussian ansatz remains appropriate. Instead, we return to Eq.~\eqref{app:E-equation} without prescribing the shape of the eigenfunctions $\tilde{h}_j$. Assuming that the first-derivative term dominates the second-derivative term gives
\begin{align}
	\partial_t \tilde{h}_j =-\frac{1}{\upsilon}\left(\frac{\lambda_j(\upsilon)}{2}+m[\cos(\omega_m t)-1]\right)\tilde{h}_j(t)+\mathcal{O}(\frac{1}{\upsilon\gamma^2}).
\end{align}
An exponential ansatz gives
\begin{align}
	\tilde{h}_j=\exp\left(\frac{2m-\lambda_j(\upsilon)}{2\upsilon}t-\frac{m}{\omega_m \upsilon}\sin(\omega_m t)\right).
\end{align}
The periodic boundary condition $\tilde{h}_j(0)=\tilde{h}_j\left(\frac{2\pi}{\omega_m}\right)$ yields
\begin{align}
	1=\exp\left(\frac{2m-\lambda_j(\upsilon)}{2\upsilon}\frac{2\pi}{\omega_m}\right)
\end{align}
This implies an eigenfunction with $\lambda_\infty(\upsilon)=2m$ that reads
\begin{align}
	\tilde{h}_\infty=e^{-\frac{m}{\omega_m\upsilon}\sin(\omega_m t)}
\end{align}
The corresponding threshold is now
\begin{align}
	G_\text{th}=k
\end{align}
Together, these limits describe a threshold that is parabolic near resonance and approaches a constant value at large detuning.

\subsubsection{Variational approach}

To describe the pulse trains found in HML and TC states, we assume Gaussian pulses with amplitudes $\rho_j$ and displacements $x_j$ from the transmission maxima $t_j=jT$. The pulse variables $(\rho_j,x_j)$ depend only on the slow time. The field ansatz is therefore
\begin{align}
	E(\theta,t)=\sum_{j=0}^{N-1}\rho_j(\theta)h_0(t-t_j-x_j(\theta))
	\label{app:eq_ansatz}
\end{align}
With the chosen normalization of $h_0$, we obtain
\begin{align}
	\int_{t_j-\sigma}^{t_j+\sigma}\vert E\vert^2 dt \sim \int_{-\infty}^\infty \vert E\vert^2dt=\rho_j^2=I_j
\end{align}
where we assume that the range $2\sigma$ is much larger than the pulse width but smaller than the period. This condition is satisfied by well-separated pulses and allows us to write
\begin{align}
	\bar{I}&=\frac{1}{\tau}\sum_j\rho_j^2=\frac{1}{\tau}\sum_j I_j\\
	\bar{G}&=\frac{G_0}{1+\bar{I}}.
\end{align}
We divide the cavity into intervals of duration $T$, each centered on a transmission maximum. We denote the gain at the beginning and end of each interval by $g_j^i$ and $g_j^f$, respectively. We shift the time origin to the center of each interval, so that the local potential minimum lies at $0$. The gain profile on $[-T/2,T/2]$ then reads
\begin{align}
	g_j(s)&=g_j^i+\bar{G}\left(\bar{I}(s+\frac{T}{2})-\int_{-T/2}^{s} I(t)dt\right)\\
	&=g_j^i+\bar{G}\left(\bar{I}(s+\frac{T}{2})-\rho_j^2\int_{-T/2}^{s} h^2(t-x_j)dt\right)
\end{align}
The integral
\begin{align}
	\eta(s)=\int_{-T/2}^{s} h_0^2(t-x_j)dt
\end{align}
increases from $0$ to $1$. Substituting the ansatz, Eq.~\eqref{app:eq_ansatz}, into Eq.~\eqref{app:E-equation} for each pulse gives
{\small
	\begin{align}
		(\partial_\theta \rho_j)&h_0(t-t_j-x_j)
		-\rho_j(\partial_\theta x_j)h_0^\prime(t-t_j-x_j)
		=\bigg(
		\frac{\bar{G}+\Gamma g-k+2m}{2}
		+\upsilon\partial_t
		+\frac{1}{2\gamma^2}\partial_t^2
		+m[\cos(\omega_m t)-1]
		\bigg)
		\rho_j h_0(t-t_j-x_j)
		\label{app:E-equation_with_ansatz}
	\end{align}
}
Introducing the local coordinate $u=t-t_j$ for the $j$-th pulse gives
{\small
	\begin{align}
		(\partial_\theta \rho_j)&h_0(u-x_j)
		-\rho_j(\partial_\theta x_j)h_0^\prime(t-t_j-x_j)
		=\left(
		\frac{\bar{G}+\Gamma g-k+2m}{2}
		+\upsilon\partial_t
		+\frac{1}{2\gamma^2}\partial_t^2
		-m\frac{\omega_m^2u^2}{2}
		\right)
		\rho_j h_0(u-x_j)\label{app:before_projection}
	\end{align}
}
The shifted mode $h_0$ satisfies
{\small
	\begin{align}
		\frac{1}{2\gamma^2}\partial_t^2 h_0(u-x_j)=\frac{-\lambda_0}{2}h_0(u-x_j)+\frac{m\omega_m^2}{2}(u-x_j)^2h_0(u-x_j).
\end{align}}
Using this identity, we obtain
{\small
	\begin{align}
		(\partial_\theta \rho_j)h_0(u-x_j)
		-\rho_j(\partial_\theta x_j)h_0^\prime(u-x_j)
		=&\big(
		\frac{\bar{G}+\Gamma g-k-\lambda_0+2m}{2}
		+\upsilon\partial_t
		+\frac{m\omega_m^2}{2}(u-x_j)^2-m\frac{\omega_m^2u^2}{2}
		\big)
		\rho_j h_0(u-x_j)\\
		=&\big(
		\frac{\bar{G}+\Gamma g-k-\lambda_0+2m}{2}
		+\upsilon\partial_t
		+\frac{m\omega_m^2}{2}(-2ux_j+x_j^2)
		\big)
		\rho_j h_0(u-x_j)
	\end{align}
} 
\subsection{Projection}
\subsubsection{Amplitude equations through projection}

We project onto the orthogonal functions $h_0$ and $h_0^\prime$ by multiplying by each function and integrating over the fast coordinate. Projection onto $h_0$ gives the amplitude equation:
\begin{align}
	\partial_\theta \rho_j=&\left(
	\frac{\bar{G}-k-\lambda_0+2m}{2}
	-\frac{m\omega_m^2}{2}x_j^2\right)\rho_j+\frac{\Gamma}{2}\rho_j\int_{-\infty}^\infty gh_0^2(u-x_j)du
\end{align}
The quadratic contribution follows from the integral
\begin{align}
	\int_{-\infty}^\infty(-2ux_j+x_j^2)h_0^2(u-x_j)du &=\int_{-\infty}^\infty(-2(z+x_j)x_j+x_j^2)h_0^2(z)dz=-x_j^2,
\end{align}
where we use the odd parity of $z h_0^2(z)$, whose integral vanishes. 

The gain contribution to the amplitude equation is
\begin{align}
	\int_{-\infty}^\infty gh_0^2(u-x_j)du=&\int_{-\infty}^\infty\left(g_j^i+\bar{G}\left[\bar{I}\left(u+\frac{T}{2}\right)-\rho_j^2\eta(u)\right]\right)h_0^2(u-x_j)du\nonumber\\
	=&g_j^i+\bar{G}\bar{I}\left(x_j+\frac{T}{2}\right)-\rho_j^2\bar{G}\int_{-\infty}^\infty\eta(u)h_0^2(u-x_j)du
\end{align}
where the last step uses the same change of variables as above. The remaining integral can be solved by noting that $\eta(s)$ can be written as $\eta^\prime+\frac{1}{2}$, where $\eta^\prime$ is an odd function that is also shifted by $x_j$. We get

\begin{align}
	\int_{-\infty}^\infty gh_0^2(u-x_j)du&=g_j^i+\bar{G}\left[\bar{I}\left(x_j+\frac{T}{2}\right)-\frac{\rho_j^2}{2}\right]
\end{align}
Combining these terms gives the evolution equation for the pulse intensity $I_j=\rho_j^2$:
\begin{align}
	\partial_\theta I_j=&\bigg[
	\bar{G}-k-\lambda_0+2m
	-m\omega_m^2x_j^2+\Gamma\left(g_j^i+\bar{G}\left[\bar{I}\left(x_j+\frac{T}{2}\right)-\frac{I_j}{2}\right]\right)\bigg]I_j\label{app:Amplitude_EOM}
\end{align}
Continuity of the gain between neighboring intervals imposes $g_{j+1}^i=g_j^f$, giving the recurrence $g_{j+1}^i=g_j^i+\bar{G}(\bar{I}T-I_j)$.

\subsubsection{Position equation}
The position equation follows from projection onto $h_0^\prime$. The integral
\begin{align}
	\int h_0^{\prime 2}=\int 4\beta^2t^2h_0^2(t)dt=\beta,
\end{align}
provides the normalization for the position equation. Projecting Eq.~\eqref{app:before_projection} gives
{\small
	\begin{align}
		-\rho_j(\partial_\theta x_j)\beta =&\rho_j\int h_0^\prime(u-x_j)\left(\upsilon\partial_t
		+\frac{m\omega_m^2}{2}(-2ux_j+x_j^2)\right) h_0(u-x_j)du+\frac{\Gamma}{2}\rho_j\int_{-\infty}^\infty g(u)h_0(u-x_j)h_0^\prime(u-x_j)du\nonumber\\
		=&\rho_j\upsilon\beta+\rho_j\frac{m\omega_m^2}{2}\int h_0^\prime(u-x_j)(-2ux_j+x_j^2) h_0(u-x_j)du+\frac{\Gamma}{2}\rho_j\int_{-\infty}^\infty g(u)h_0(u-x_j)h_0^\prime(u-x_j)du
\end{align}}
Using the same change of variables as above, the first integral becomes
\begin{align}
	\int h_0^\prime(u-x_j)(-2ux_j+x_j^2) h(u-x_j)du=&\int h_0^\prime(z)(-2zx_j-x_j^2) h(z)dz\\
	=&-x_j\int v 2h_0^\prime(z) h_0(z)dz\\
	=&-x_j\left([zh_0^2]^\infty_{-\infty}-\int h_0^2(z) dz\right)=x_j,
\end{align}
This contribution produces a restoring force toward the potential minimum. The gain contribution involves the second integral:
{\small
	\begin{align}
		\int_{-\infty}^\infty g(u)h_0(u-x_j)h_0^\prime(u-x_j)du=&\int_{-\infty}^\infty \left(g_j^i+\bar{G}\left[\bar{I}\left(u+\frac{T}{2}\right)-\rho_j^2\eta(u)\right]\right)h_0(u-x_j)h_0^\prime(u-x_j)du\nonumber\\
		=&\frac{\bar{G}}{2}\int_{-\infty}^\infty \left[\bar{I}\left(u+\frac{T}{2}\right)-\rho_j^2\eta(u)\right]2h_0(u-x_j)h_0^\prime(u-x_j)du\nonumber\\
		=&\frac{\bar{G}}{2} \left[\left(\bar{I}\left(u+\frac{T}{2}\right)-\rho_j^2\eta(u)\right)h_0^2(u-x_j)\right]^\infty_{-\infty}-\frac{\bar{G}}{2}\int_{-\infty}^\infty \left[\bar{I}-\rho_j^2h_0^2(u-x_j)\right]h_0^2(u-x_j)du\nonumber\\
		=&-\frac{\bar{G}}{2}\left(\bar{I}-\rho_j^2\int_{-\infty}^\infty h_0^4(u-x_j)du\right)\nonumber\\
		=&-\frac{\bar{G}}{2}\left(\bar{I}-\rho_j^2\sqrt{\frac{\beta}{\pi}}\right)\,.
\end{align}}
In summary, we find
\begin{align}
	-\rho_j(\partial_\theta x_j)\beta =&\rho_j\upsilon\beta+\rho_j\frac{m\omega_m^2}{2}x_j-\Gamma\frac{\bar{G}}{4}\left(\bar{I}-\rho_j^2\sqrt{\frac{\beta}{\pi}}\right)\rho_j
\end{align}
which yields the position equation
\begin{align}
	\partial_\theta x_j =&-\upsilon-\frac{m\omega_m^2}{2\beta}x_j+\Gamma\frac{\bar{G}}{4\beta}\left(\bar{I}-I_j\sqrt{\frac{\beta}{\pi}}\right)\label{app:Position_EOM}
\end{align}

\subsubsection{Gain equation}
The amplitude and position equations are supplemented by the gain recurrence $g_{j+1}^i=g_j^i+\bar{G}(\bar{I}T-I_j)$. To close the system, we must determine the gain $g_1^i$ at the beginning of the first interval. We impose the zero-mean condition on $g$ using Eq.~\eqref{app:g2_equation}:
\begin{align}
	g(\theta,t)&=\bar{G}\left(\bar{I}(t-\frac{\tau}{2})-P(t)+\bar{P}\right)\\
	P(t)&=\int_0^t I(\theta,s)ds.
\end{align}
This yields
\begin{align}
	g^i=g(\theta,0)&=\bar{G}\left(\bar{P}-\frac{\tau}{2}\bar{I}\right)=\bar{G}\left(\bar{P}-\frac{1}{2}\sum_j I_j\right),
\end{align}
It therefore remains to evaluate $\bar{P}$, the average cumulative intensity over one round trip. In the narrow-pulse approximation, each pulse contributes a jump at $\frac{(2j-1)T}{2}+x_j$. Summing these contributions gives
\begin{align}
	\bar{P}=&\frac{1}{\tau}\bigg(I_1\left[\tau-\left(\frac{T}{2}+x_1\right)\right]+I_2\left[\tau-\left(\frac{3T}{2}+x_2\right)\right]+\cdots+I_N\left[\tau-\left(\frac{(2N-1)T}{2}+x_N\right)\right]\bigg) \\
	=&\frac{1}{\tau}\sum_{j=1}^N I_j\left(\frac{(2N+1-2j)T}{2}-x_j\right)=\frac{1}{NT}\sum_{j=1}^N I_j\left(\frac{(2N+1-2j)T}{2}-x_j\right).
\end{align}
Substituting this expression into the initial condition gives
\begin{align}
	g_1^i&=\bar{G}\sum_{j=1}^N I_j\left(\frac{N+1-2j}{2N}-\frac{x_j}{\tau}\right)\,.
\end{align}

\subsection{Effective equations of motion}
Using $\beta=\gamma\omega_m\sqrt{m}/2$ and $\lambda_0=(\omega_m/\gamma)\sqrt{m}$, so that $\frac{m\omega_m^2}{2\beta}=\lambda_0$, we obtain
\begin{align}
	\partial_\theta I_j=&\bigg[
	\bar{G}-k-\lambda_0+2m
	-m\omega_m^2x_j^2+\Gamma\left(g_j^i+\bar{G}\left[\bar{I}\left(x_j+\frac{T}{2}\right)-\frac{I_j}{2}\right]\right)\bigg]I_j\\
	\partial_\theta x_j =&-\upsilon-\lambda_0 x_j+\Gamma\frac{\bar{G}}{4\beta}\left(\bar{I}-I_j\sqrt{\frac{\beta}{\pi}}\right)\\
	g_1=&\bar{G}\sum_{j=1}^N I_j\left(\frac{N+1-2j}{2N}-\frac{x_j}{\tau}\right)\\
	g_{j+1}=&g_j+\bar{G}(\bar{I}T-I_j).
\end{align}
The relations $\bar{G}=\frac{G_0}{1+\bar{I}}$ and $\bar{I}=\frac{1}{\tau}\sum_j I_j$ complete the system. We now introduce normalized variables to express it more compactly.

\subsubsection{Normalization}
We first scale the gain and pump parameter by the effective losses $l=k+\lambda_0-2m$. Introducing $\bar{G}/l \rightarrow\tilde{G}$, $g/l \rightarrow\tilde{g}$, $G_0/l \rightarrow J$, $\frac{1}{l} \partial_\theta \rightarrow \partial_{\tilde{\theta}}$ gives
\begin{align}
	\partial_{\tilde{\theta}} I_j=&\bigg[
	\tilde{G}-1
	-\frac{m\omega_m^2}{l}x_j^2+\Gamma\left(\tilde{g}_j^i+\tilde{G}\left[\bar{I}\left(x_j+\frac{T}{2}\right)-\frac{I_j}{2}\right]\right)\bigg]I_j\\
	\partial_{\tilde{\theta}} x_j =&-\frac{\upsilon}{l}-\frac{\lambda_0}{l}x_j+\Gamma\frac{\tilde{G}}{4\beta}\left(\bar{I}-I_j\sqrt{\frac{\beta}{\pi}}\right)\\
	\tilde{g}_1=&\tilde{G}\sum_{j=1}^N I_j\left(\frac{N+1-2j}{2N}-\frac{x_j}{\tau}\right)\\
	\tilde{g}_{j+1}=&\tilde{g}_j+\tilde{G}(\bar{I}T-I_j)
\end{align}
with $\tilde{G}=\frac{\tilde{J}}{1+\bar{I}}$. Next, using $\tau=NT$, we introduce $\frac{x_j}{T}\rightarrow y_j$, $\braket{I}=\frac{1}{N}\sum_j I_j$ so that $\bar{I}=\frac{1}{\tau}\sum_j I=\braket{I}/T$. The equations become
\begin{align}
	\partial_{\tilde{\theta}} I_j=&\bigg[
	\tilde{G}-1
	-\frac{m\omega_m^2 T^2}{l}y_j^2+\Gamma\left(\tilde{g}_j^i+\tilde{G}\left[\braket{I}\left(y_j+\frac{1}{2}\right)-\frac{I_j}{2}\right]\right)\bigg]I_j\\
	\partial_{\tilde{\theta}} y_j =&-\frac{\upsilon}{l T}-\frac{\lambda_0}{l}y_j+\Gamma\frac{\tilde{G}}{4\beta T}\left(\frac{\braket{I}}{T}-I_j\sqrt{\frac{\beta}{\pi}}\right)\\
	\tilde{g}_1=&\tilde{G}\sum_{j=1}^N I_j\frac{N+1-2j-2y_j}{2N}\\
	\tilde{g}_{j+1}=&\tilde{g}_j+\tilde{G}(\braket{I}-I_j)
\end{align}
with $\tilde{G}=\frac{J}{1+\braket{I}/T}$. We then scale the intensity as $S=I/T$ and the gain fluctuation as $ \tilde{g}_j/T \rightarrow \hat{g}_j$, and define $\Gamma T \rightarrow \varrho $. Using $\omega_m T=2\pi$ and $\beta=\gamma\omega_m\sqrt{m}/2$ gives
{\small
	\begin{align}
		\partial_{\tilde{\theta}} S_j=&\bigg[
		\tilde{G}-1
		-\frac{4 \pi^2 m}{l}y_j^2+\varrho\left(\hat{g}_j^i+\tilde{G}\left[\braket{S}\left(y_j+\frac{1}{2}\right)-\frac{S_j}{2}\right]\right)\bigg]S_j\\
		\frac{l}{\lambda_0}\partial_{\tilde{\theta}} y_j =&-\frac{\upsilon}{\lambda_0 T}-y_j+\frac{\varrho\tilde{G}l}{8\pi^2m}\left(\braket{S}-S_jT\sqrt{\frac{\beta}{\pi}}\right)\\
		\hat{g}_1=&\tilde{G}\sum_{j=1}^N S_j\frac{N+1-2j-2y_j}{2N}\\
		\hat{g}_{j+1}=&\hat{g}_j+\tilde{G}(\braket{S}-S_j)
\end{align}}
with $\tilde{G}=\frac{J}{1+\braket{S}}$. Finally, defining $\tilde{\upsilon}=\frac{\upsilon}{\lambda_0 T}$, $\frac{l}{\lambda_0}=\mu^{-1}$ and $r=T\sqrt{\beta/\pi}=\sqrt{T\gamma\sqrt{m}}$, gives $4\frac{\pi^2 m}{l}=2\pi\mu r^2$ and the final system
\begin{align}
	\partial_{\tilde{\theta}} S_j=&\bigg[
	\tilde{G}-1
	-2\pi\mu r^2 y_j^2+\varrho\left(\hat{g}_j^i+\tilde{G}\left[\braket{S}\left(y_j+\frac{1}{2}\right)-\frac{S_j}{2}\right]\right)\bigg]S_j \label{app:eq:EOM1}\\
	\frac{1}{\mu}\partial_{\tilde{\theta}} y_j =&-\tilde{\upsilon}-y_j+\frac{\varrho}{4\pi\mu r^2}\tilde{G}\left(\braket{S}-S_jr\right) \label{app:eq:EOM2}\\
	\hat{g}_1=&\tilde{G}\sum_{j=1}^N S_j\frac{N+1-2j-2y_j}{2N}\label{app:eq:EOM3}\\
	\hat{g}_{j+1}=&\hat{g}_j+\tilde{G}(\braket{S}-S_j)\label{app:eq:EOM4}
\end{align}
with $\tilde{G}=\frac{J}{1+\braket{S}}$.

\clearpage
\section{Additional bifurcation analysis}
\label{app:bifurcation}
\begin{figure}[htbp]
    \centering
    \includegraphics[width=\linewidth,height=0.64\textheight,keepaspectratio]{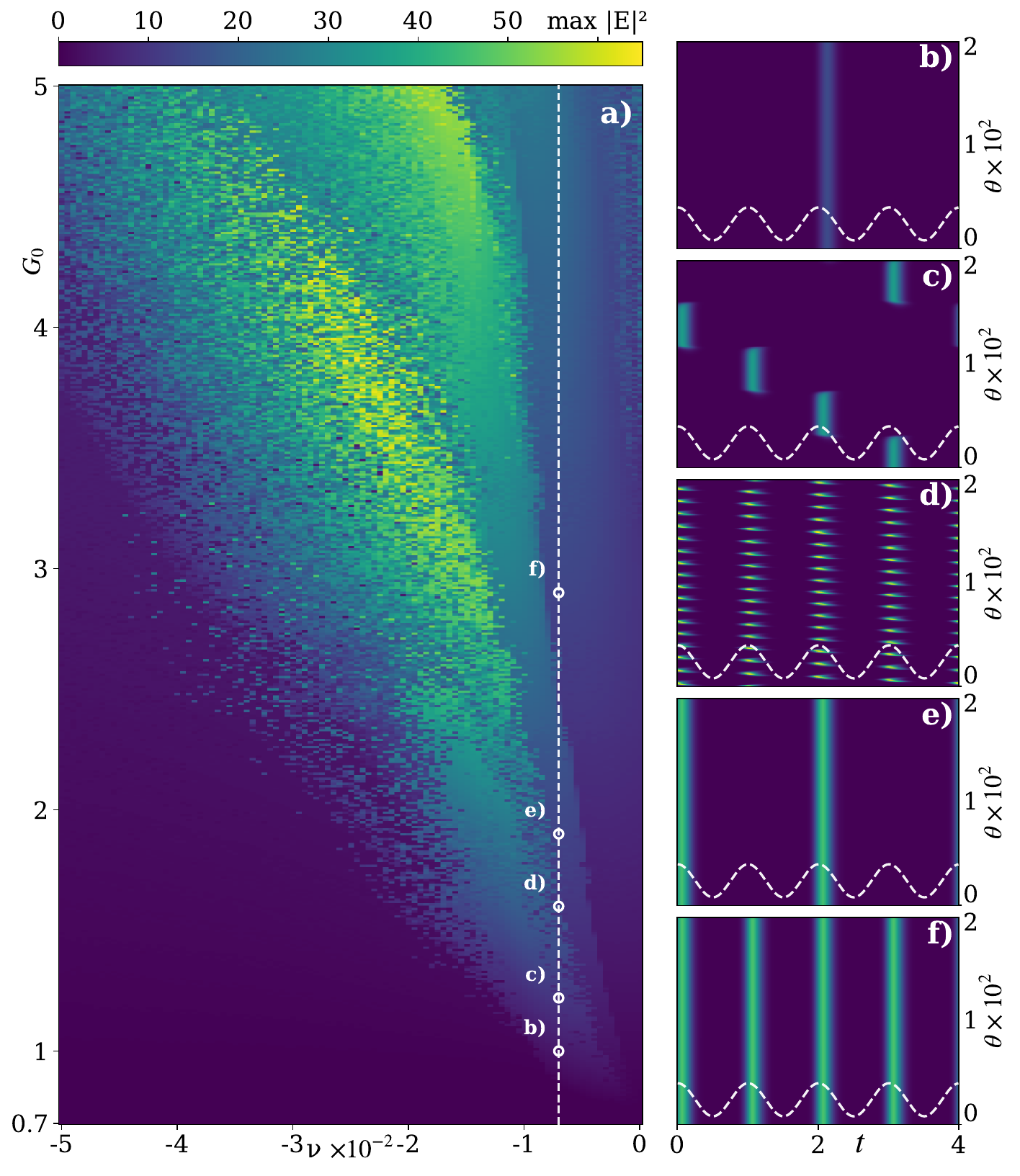}
    \caption{(a) Bifurcation diagram in the $(\upsilon,\,G_0)$ plane obtained with direct numerical simulations of the HME model given by Eqs.~\eqref{eq:HME1}--\eqref{eq:BC_HME} for $\alpha=1.5$. The colormap corresponds to $\max(|E|^2)$. (b)--(f) Exemplary field intensities in a two-time representation calculated at points along the white line at $\upsilon=-0.007$. An excerpt of the last $200$ round trips of a long-time simulation over $2\times10^5$ round trips is presented. Other parameters are as in Fig.~\ref{fig:dns_2param}.}
    \label{app:fig:fig:dns_2param_alp}
\end{figure}

Figure~\ref{app:fig:fig:dns_2param_alp} presents a numerical bifurcation diagram in the $(\upsilon,\,G_0)$ plane obtained for the more realistic value $\alpha=1.5$, using the HME model given by Eqs.~\eqref{eq:HME1}--\eqref{eq:BC_HME}.
Here, a split-step method with $N_x=256$ grid points on the domain $t\in(0,\,4)$ was employed for the numerical integration. The initial condition was the off state perturbed by weak noise. The colormap represents the maximum field intensity. The vertical white line marks the cut through the plane at $\upsilon=-0.007$. The labeled points along this line correspond to the $E$-field intensity profiles shown in a two-time representation in panels (b)--(f). These illustrate the stable TC$_4$ (panel (b)), oscillatory TC$_2$ (panel (c)), oscillatory HML$_4$ (panel (d)), stable TC$_2$ (panel (e)), and stable HML$_4$ (panel (f)) regimes, respectively.

\end{document}